\documentclass[12pt,graphics]{iopart}

\def\RR{\hbox{{\rm I}\kern-.2em\hbox{\rm R}}}
\def\pRR{\hbox{{\tiny \rm I}\kern-.1em\hbox{{\tiny \rm R}}}}

\def\NN{\hbox{I\kern-.2em\hbox{N}}}

\input{epsf}

\begin{document}

\title{Exact solutions and dynamics of globally coupled phase
oscillators}
\author{Luis L. Bonilla \footnote{To whom correspondence should be
addressed}}
\address{Departamento de Matem\'aticas. Escuela Polit\'ecnica
Superior,  Universidad Carlos III de Madrid,\\
Avenida de la Universidad 30.  28911 Legan{\'e}s, Spain.  e-mail:
{\tt bonilla@ing.uc3m.es}}
\author{Conrad J. P\'erez Vicente \& Felix Ritort}
\address{Departament de Fisica Fonamental, Universitat de
Barcelona, Diagonal 647  \\ 08028 Barcelona, Spain. email: {\tt
conrad@hermes.ffn.ub.es},  {\tt ritort@ffn.ub.es}}
\author{Juan Soler }
\address{Departamento de Matem\'atica Aplicada, Facultad de Ciencias,
Universidad de Granada \\18071 Granada, Spain. email: {\tt
jsoler@ugr.es}}

\date{\today}

\maketitle

\begin{abstract} We analyze mean-field models of
synchronization of phase oscillators
with singular couplings and subject to external
random forces. They are related to the
Kuramoto-Sakaguchi model. Their probability
densities satisfy local partial differential
equations similar to the Porous Medium, Burgers
and extended Burgers equations depending on the
degree of singularity of the coupling. We show
that Porous Medium oscillators (the most
singularly coupled) do not synchronize and that
(transient) synchronization is possible only at
zero temperature for Burgers oscillators. The
extended Burgers oscillators have a nonlocal
coupling first introduced by Daido and they may
synchronize at any temperature. Exact expressions
for their synchronized phases and for Daido's
order function are given in terms of elliptic
functions.
\end{abstract}

\vskip.5pc




\section{Introduction}
\label{sec:1}
Collective synchronization of large populations
of nonlinearly coupled phase oscillators has been
intensely studied since Winfree realized its
importance for biological systems~\cite{winfree}
and Kuramoto gave  mathematical form to these
ideas in a simple
model~\cite{kuramoto,kuramoto2,kuramoto3}.
Motivation for studying the Kuramoto model can be
found in the broad  variety of physical, chemical
or biological phenomena which can be modelled
within its framework (see
\cite{winfree,Ku84,Stro94,wiesen} and references
therein).


The problem we want to consider is the dynamics of a system of
nonlinearly,  globally coupled phase oscillators with random
frequencies $\omega_i$ [the  probability for an oscillator to have
frequency $\omega_j \in (\omega,\omega+ d\omega)$ is $p(\omega)\,
d\omega$], subject to external (independent,
identically-distributed) white noise sources $\eta_{i}$ (of
strength $\sqrt{2T}$):
\begin{equation}
\frac{\partial\phi_i}{\partial t}= \omega_{i} -
\frac{K_0}{N}\sum_{j=1}^N f(\phi_i-\phi_j) + \eta_i(t),
\label{eq1}
\end{equation}
$i=1,\ldots,N$. Here $\phi_{i}(t)$ denotes the {\it i}th oscillator
phase,
$K>0$ represents the coupling strength, and $f$ is a generic real
function of periodicity $2\pi$. In Fourier space, the latter can be
decomposed as follows,
\begin{equation}
f(x)=\sum_{n=-\infty}^{\infty} a_n \exp(inx)
\label{eq2}
\end{equation} with $a_n=a_{-n}^*$.

Important work on the Kuramoto model with a
general coupling function was carried out by
Daido~\cite{daidoOF,daido2} and
Crawford~\cite{Cr} among others. Daido introduced
the concept of order function  to understand
synchronization at zero temperature, $T=0$.
Essentially, the order function is an average of
the oscillator drift velocity on a rotating
frame~\cite{daido,daido2}. Assuming a reasonable
shape for the order function and  the
one-oscillator probability density (supposedly
stationary in the rotating  frame), Daido derived
a functional equation for the order function. He
then  found solutions by direct numerical
simulations and by bifurcation theory for
stationary or rotating-wave probability
densities. However, his theory is not
sufficiently general to cover other possible
probability densities (e.g., standing
waves~\cite{Cr94,BPS}) nor to predict their
stability properties. Crawford considered the
problem of constructing one-oscillator
probability  densities which bifurcate from the
incoherent density $1/(2\pi)$ (for which an
oscillator has the same probability to be at any
given value of the angle $\phi$)~\cite{Cr}. As a
consequence of these works, different conjectures
were  proposed, both concerning scaling of
bifurcating solution branches and general
statements such as ``adding noise could actually
yield an enhancement in the  level of
synchronization'' \cite{Cr}. One difficulty with
these previous works is that explicit
calculations were performed mostly near
bifurcation points. One exception is Daido's
coupling function $f(x) = $ sign$x$ for which the
calculation of the order function can be
explicitly carried out.

In this paper, we shall introduce classes of
coupling functions leading to  drift terms which
are {\em local} functions of the probability
density. We restrict ourselves to oscillators
without frequency disorder, i.e.,  with frequency
distribution $p(\omega)=$ $\delta(\omega)$. The
corresponding governing equations for the
probability density (in the limit of infinitely
many oscillators) are systems of nonlinear
partial differential equations which are
analyzed. Then definite results can be proved
about synchronization which do not depend on
bifurcation calculations, order function theory
or numerical simulations as much as the  work of
previous authors. Together with singular
perturbation calculations, our results can be
used to give approximations to the one-oscillator
probability density of models with disorder, in
the limiting case of high-frequencies \cite{AB98}.
The examples of singular coupling functions
considered in this paper are the hard-needles  (+
sign) and stick-needles (-- sign) couplings, for
which $f = \pm\delta'$, the Burgers coupling, $f
= \delta$, and the Daido coupling (extended
Burgers model), $f(x) =$ sign$(x)$. In all cases,
these $f$'s are extended periodically outside the
interval $-\pi<x<\pi$. We shall show that less
singular couplings give  rise to models for which
the oscillators are easier to synchronize:  (i)
the hard-needles oscillators do not synchronize
at any temperature (the stick-needles coupling
leads to ill-posed problems unless the
temperature is large enough), (ii) the Burgers
oscillators may synchronize only at zero
temperature; even then, their probability density
algebraically decays to incoherence as $t^{-1}$
for large times, (iii) the oscillators with Daido
coupling may synchronize at any temperature, and
the synchronized phases are described by
explicitly-known order functions and probability
densities. A first paper in the direction of
these results was published by the authors in
\cite{BPRS}.

The rest of the paper is structured as follows.
Section \ref{sec-general} contains results which
hold for general models with disorder in the
natural frequencies. These include a derivation of
the governing equations for the one-oscillator
probability  density, its moments and the
moment-generating function. We establish a
relationship between these objects and Daido's
order function showing that the latter is
proportional to the average oscillator drift in a
rotating frame. Lastly, we give the leading-order
form of the probability density in the limit of
high frequencies. The density is a superposition
of probability densities with zero natural
frequency in rotating frames and with smaller
coupling constants. In Section \ref{sec-odd} we
extend Daido's approach to the case of nonzero
temperature, and find the form of the stationary
solutions and a general Liapunov functional for
the case of odd coupling functions. In section
\ref{sec-family} we introduce the family of
singular couplings to be studied. Section
\ref{sec-porous} contains our analysis of the
porous Medium models which appear for the
hard-needles and stick-needles couplings. We show
that the hard-needles model is well-posed, its
solution exists globally, and also that sharp
time decay estimates towards its equilibrium can
be obtained locally for any initial data or
globally for small initial data  evolving
exponentially fast towards incoherence. The last
result is proven by different methods according
to whether we allow the temperature to be zero.
On the other hand, the stick-needles model is
ill-posed for low enough temperature. Section
\ref{sec-burgers} is devoted to study a model
with Burgers coupling. We adapt well-known
results to the case of periodic boundary
conditions to show that the probability density
of the Burgers  oscillators may tend to a state
different from incoherence only if the
temperature is zero. In section \ref{sec-daido}
we analyze the extended Burgers model resulting
from the Daido coupling $f(x) = $ sign$(x)$. We
first show that finding the probability density
is equivalent to solving two coupled nonlinear
parabolic equations for the drift $v(x,t)$ and a
certain functional of the probability density,
$\sigma(x,t)$. Then we find families of
stationary solutions in terms of elliptic
functions. These solutions bifurcate
supercritically from incoherence at  couplings
$K_0$ which are proportional to the squares of
odd integer numbers, and stationary densities on
the first bifurcating branch are stable at least
for small enough $K_0$. We discuss some of the
results of previous authors  in the light of our
exact calculations. The Appendices discuss
technical matters related to the bulk of the
article.


\section{Probability density and moment-generating
function}
\label{sec-general}

In this section we consider general aspects
of the model (\ref{eq1}). First of all, we use the
moment approach considered in \cite{PR} to derive
a nonlinear Fokker-Planck equation (NLFPE) for
the one-oscillator probability density in the
limit of infinitely many oscillators. The moment
approach exploits the symmetry of the dynamical
problem we are interested in, so is a good
starting point to deal with the Kuramoto model
which has rotational symmetry (an extension of
this method to deal with tops has been considered
in \cite{RITORT}). Secondly, we characterize
phase and frequency synchronization generalizing
to $T\neq 0$ the concept of order function
introduced by Daido \cite{daidoOF} for oscillator
synchronization at $T=0$. Lastly, we show that, at
high frequencies, the probability density can be
decomposed into $m$ components rotating steadily
at the frequencies of the peaks of a given
multimodal natural frequency distribution. Each
component probability density solves a NLFPE with
zero natural frequency (in the rotating frame)
and a modified coupling constant. This latter
results follows immediately from the method
introduced in Ref.\ \cite{AB98} for the usual
Kuramoto model with a sinusoidal coupling
function. This result can be combined with the
exact solutions obtained in later Sections to
yield analytic expressions of synchronized phases
in models with frequency disorder and white noise
forcing {\em in the high-frequency limit}.

\subsection{Nonlinear Fokker-Planck equation}
To start with let us  define,
\begin{equation}
H_{k}^m=\frac{1}{N}\sum_{j=1}^{N}\,\overline{
\langle
\exp(ik\phi_j)\rangle \omega_j^m}\,,
\label{eq3}
\end{equation}
where the brackets denote average with respect to the
external noise and the overbar average with respect to the random
oscillator frequency. This set of moments is invariant under the local
symmetry  $\phi_i\to \phi_i+2\pi$ which is the symmetry of the
dynamical equations \ref{eq1}. The equation of motion for the moments reads
\begin{eqnarray}
\frac{\partial H_{k}^m}{\partial t}=-K_0 i k
\sum_{n=-\infty}^{\infty} a_n H_{k+n}^m H_{-n}^0  - k^2 T H_{k}^m\,
+ \,ik H_{k}^{m+1} .
\label{eq4}
\end{eqnarray}
It is easy to derive from Eq.(\ref{eq4}) the
following differential equation
\begin{equation}
\frac{\partial g}{\partial t} = -\frac{\partial}{\partial x}
\Bigl [ v(x,t)\, g \Bigr ] + T\frac{\partial^2 g}{\partial
x^2}-\frac{\partial^2 g }{\partial x\partial y} \,
\label{eq6}
\end{equation}
for the generating function
\begin{equation} g(x,y,t)= {1\over 2\pi}\,
\sum_{k=-\infty}^{\infty}\sum_{m=0}^{\infty}
\exp(-ikx)\frac{y^m}{m!}H_k^m(t) .
\label{eq5}
\end{equation}
The velocity $v(x,t)$ is defined by
\begin{equation} v(x,t)=-K_0 \sum_{n=-\infty}^{\infty}a_n
H_{-n}^0\exp(inx)
\label{eq7}
\end{equation}
and it can be equivalently written as a
convolution between the coupling and the
generating function (see \ref{app:A}):
\begin{equation}
v(x,t) \ = \ -K_0\,\int_{-\pi}^{\pi} f(x')
g(x-x',0,t) \ dx' .
\label{conv}
\end{equation}
Notice that this drift velocity is related to Daido's
order function
$H(x,t)$~\cite{daido} (see \ref{app:B}):
\begin{equation}
v(x,t) = - K_0 \, H(x-\Omega_e t,t).
\label{OF}
\end{equation}

The case analyzed by Kuramoto corresponds to the
force $f(x)=\sin x$. Then $a_1=a_{-1}^*=-i/2$ and
the rest of components are zero. This yields
$v(x,t) = K_0 r \sin(\theta-x)$, where $H_1^0=
r\exp(i\theta)$.  Let us assume that there exists
a probability density $\rho(x,\omega,t)$ such that
\begin{eqnarray}
{1\over N}\sum_{j=1}^{N} {\cal
F}(\phi_j(t),\omega_j) = \int_{-\pi}^{\pi}
\int_{-\infty}^{\infty} {\cal F}(x,\omega)\,
\rho(x,\omega,t)\, p(\omega)\, d\omega\, dx,
\label{limit1}
\end{eqnarray}
as $N\to\infty$. In particular, we assume that

\begin{eqnarray}
\rho(x,\omega,t)\, p(\omega) = {1\over N}\sum_{j=1}^{N}
\delta(x-\phi_j(t))\, \delta(\omega-\omega_j)\,,\label{limit2}
\end{eqnarray}
tends to a smooth function in the limit as
$N\to\infty$. We can relate the probability
density to the moment-generating function as
follows:
\begin{eqnarray}
g(x,y,t) = {1\over 2\pi}\sum_{k=-\infty}^{\infty}
\sum_{m=0}^{\infty} e^{-ikx} {y^{m}\over
m!}{1\over N}\sum_{j=1}^{N}
e^{ik\phi_{j}(t)}\omega_j^m \nonumber\\
= {1\over N}\sum_{j=1}^{N} e^{y\omega_{j}} {1\over
2\pi}\sum_{k=-\infty}^{\infty} e^{ik[\phi_{j}(t) -
x]}  = {1\over N}\sum_{j=1}^{N} e^{y\omega_{j}}
\delta(\phi_{j}(t) - x)\nonumber\\
= \int_{-\infty}^{\infty}
e^{y\omega}\,\rho(x,\omega,t)\, p(\omega)\,
d\omega.
\label{limit3}
\end{eqnarray}
Then the probability density can be written as
follows
\begin{eqnarray}
\rho(x,\omega,t)\, p(\omega) = {1\over
2\pi}\int_{-\infty}^{\infty} g(x,iy,t) \,
e^{-i\omega y}\, dy
\label{limit4}
\end{eqnarray}
and
\begin{eqnarray}
-{1\over 2\pi i}\int_{-\infty}^{\infty}
e^{-i\omega
y} {\partial g\over\partial  y}\, dy =
-\omega\,\rho(x,\omega,t)\,
p(\omega). \label{limit5}
\end{eqnarray}
By using these expressions in (\ref{eq6}), we derive
the usual nonlinear  Fokker-Planck equation
(NLFPE) for $\rho(x,\omega,t)$ wherever
$p(\omega)\neq 0$:
\begin{equation}
\frac{\partial \rho}{\partial t} + \frac{\partial}{\partial x}
\Bigl [ (\omega + v)\, \rho \Bigr ] = T\frac{\partial^2
\rho}{\partial x^2} \,. \label{limit6}
\end{equation}
Equations (\ref{limit1}) and (\ref{conv}) provide
us with the following formula  for the drift
velocity $v$ in terms of $\rho$:
\begin{eqnarray}
v(x,t) \ = \
 -  K_0 \int_{-\pi}^{\pi}\int_{-\infty}^{\infty} f(x')
\rho(x-x',\omega,t) p(\omega)\, dx' d\omega . \label{limit7}
\end{eqnarray}
The probability density $\rho$ should moreover
satisfy a normalization condition
$\int_{-\pi}^{\pi}
\rho dx = 1$ and an initial condition
$\rho(x,\omega,0)=\rho_0(x,\omega)$. A derivation
of these problems by path integral methods can be
found in \cite{bon2}. (The path integral
derivation is applicable to each Fourier mode of
the  coupling function; see page 676 of
\cite{bon2}). See \cite{dawson,daipra} for
rigorous proofs of (\ref{limit1}), (\ref{limit6})
- (\ref{limit7}) in different models.

\subsection{Phase and frequency distributions}
To characterize oscillator synchronization, it is convenient to
define  the phase and frequency probability densities. The phase
probability  density, $ P(\phi,t)$, is the probability of finding
an oscillator with angle in $(\phi,\phi+d\phi)$ independently of
its frequency. It is given by
\begin{eqnarray}
 P(\phi,t) =
\int_{-\pi}^{\pi}\int_{-\infty}^{\infty} \delta(x-\phi)\,
\rho(x,\omega,t)\, p(\omega)\, dx\, d\omega . \label{limit8}
\end{eqnarray}
Similarly, we may define the frequency of a given
oscillator as the average  (when it exists)
$$ {1\over\tau}\, \int_0^\tau {d\phi_{j}\over dt}\, dt ,$$ for
sufficiently large $\tau >0$. When $N\to\infty$, (\ref{eq1}) and the
ergodic theorem imply that the previous time average is equal to
$$ \omega + {1\over\tau}\, \int_0^\tau v(x,t)\, dt ,$$ where $\omega
= \omega_j$. Then we may define the frequency density as
\begin{eqnarray}
P(\omega,t) = \int_{-\pi}^{\pi}
\int_{-\infty}^{\infty}
\delta\left(\Omega-\omega - {1\over\tau}\,
\int_0^\tau v(x,t')\, dt'\right)
\rho(x,\Omega,t)\, p(\omega)\, dx\, d\Omega
\label{limit9}
\end{eqnarray}
that can be simplified in the important
cases of:  stationary (i)  and  rotating-wave (ii) probability
densities. In case (i), $v=v(x)$ coincides  with its time average,
and is proportional to Daido's order function with
$\Omega_e = 0$. In case (ii), $\rho(x,\Omega,t) =
\tilde{\rho}(x-\Omega_e t,
\Delta)$, with $\Delta = \Omega -\Omega_e$ and, according
(\ref{OF}), $v(x,t) = - K_0\, H(x-\Omega_e t)$. Then the change of
variable $\psi = x-\Omega_e t$ reduces  probability density and
order function to time-independent functions so that  (\ref{limit9})
becomes
 \begin{eqnarray}
P(\omega) =
\int_{-\pi}^{\pi}\int_{-\infty}^{\infty} \delta[\omega-\Omega_e -
\Delta + K_0\, H(\psi)]\, \tilde{\rho}(\psi,\Delta)\,
\tilde{p}(\Delta)\, d\psi\, d\Delta , \label{limit9.stat}
\end{eqnarray}
 where $\tilde{p}(\Delta) = p(\Omega_e + \Delta)$.
Daido considered only cases  (i) and (ii) at $T=0$ with an order
function having a single minimum $H_{min}$  (resp.\ maximum
$H_{max}$) at $\psi = \psi_{1}$ (resp.\ $\psi = \psi_{2}$)
\cite{daidoOF}.
Then the probability density is
\begin{eqnarray}
\tilde{\rho}(\psi,\Delta) = \delta\left(\psi -
H^{-1}\left({\Delta\over K_{0}}
\right)\right)\, \chi_{(\psi_{1},\psi_{2})}(\psi)\nonumber\\
 + {C(\Delta)\over \Delta - K_{0}\, H(\psi)}\, \left[
1-\chi_{(\psi_{1},
\psi_{2})}(\psi)\right]\,,
\label{limit10}
\end{eqnarray}
where
\begin{eqnarray}
C(\Delta) = {2\pi\over\int_{-\pi}^{\pi}
{d\psi\over\Delta-K_{0}H(\psi)} }\,,
\label{limit11}
\end{eqnarray}
$ \psi = \phi - \Omega_e t$, with $\Delta$ as before and
$\chi_{(\psi_{1},\psi_{2})}(\psi)$ equals 1 if
$\psi_{1}<\psi<\psi_{2}$ and  0 otherwise. On the interval
$-\pi<\psi<\pi$ the order function $H(\psi)$   is stationary as
before. $C(\Delta)$ is the frequency at which oscillators  with
angle outside $(\psi_1,\psi_2)$ rotate. Inside $(\psi_1,\psi_2)$ we
have
$K_{0} H(\psi)=\Delta$ and $d\psi/dt = 0$. Inserting (\ref{limit10})
in (\ref{limit7}) and (\ref{OF}) we find a functional equation for
the order function, which  can then be solved exactly or
approximately~\cite{daidoOF}. Daido's expressions  for the angle and
frequency densities are found by inserting (\ref{limit10}) in
(\ref{limit8}) and (\ref{limit9.stat}) (minor notational changes
have been  made)~\cite{daidoOF}
\begin{eqnarray}
 P(\psi) = K_0\, \tilde{p}(K_0 H(\psi))\,
H'(\psi)\,\chi_{(\psi_{1},
\psi_{2})}(\psi)
 + \left[1-\chi_{(\psi_{1},\psi_{2})}(\psi)\right]\,
\nonumber\\ \times \int_{-\infty}^{\infty}
{C(\Delta)\tilde{p}(\Delta)\over\Delta - K_{0}\, H(\psi)}\,
\chi_{(-\infty,K_{0} H_{min})}(\Delta)\, \chi_{(K_{0}
H_{max},\infty)}(\Delta)\, d\Delta , \label{limit12}\\ P(\omega) =
\delta(\omega-\Omega_e)\, \int_{K_{0}H_{min}}^{K_{0}H_{max}}
\tilde{p}(\Delta) d\Delta  + {\tilde{p}(C^{-1}(\omega-\Omega_{e}))
\over C'(C^{-1}(\omega-\Omega_{e}))}\,. \label{limit13}
\end{eqnarray}

\subsection{High-frequency limit }
Further general considerations can be made for
multimodal natural frequency distributions in
the limit of high frequencies \cite{AB98}. Let us
assume that $p(\omega)$ has $m$ maxima located at
$\omega_0\Omega_l$, $l=1,\ldots,m$, where
$\omega_0\rightarrow \infty$, and $p(\omega)\,
d\omega$ tends to the limit distribution
\begin{eqnarray}
\Gamma(\Omega)\, d\Omega \equiv \sum_{l=1}^{m}
\alpha_l\,\delta(\Omega  -\Omega_l)\, d\Omega,
\label{h1}\\
\mbox{with}\quad\quad \sum_{l=1}^{m} \alpha_l =
1,\quad\mbox{and}\quad
\Omega = \frac{\omega}{\omega_{0}}\,, \nonumber
\end{eqnarray}
independently of the shape of $p(\omega)$ as
$\omega_0\to\infty$. $p(\omega)\, d\omega$ and
$\Gamma(\Omega)\, d\Omega$ may be used
interchangeably when calculating any moment of
the probability density [including of course the
all-important velocity function (\ref{limit7}),
which is related to Daido's order function as
said before]. Thus any frequency distribution is
equivalent to a discrete multimodal distribution
in the high-frequency limit. The discrete
symmetric bimodal distribution  considered in
\cite{BNS,BPVS} corresponds to $m=2$, $\Omega_l =
(-1)^l$, $\alpha_l = {1\over 2}$, $l=1,2$. By
following the procedure explained in \cite{AB98},
we can show that the oscillator probability
density splits into $m$ components, each
contributing a wave rotating with frequency
$\Omega_l\omega_0$ to the order function:
\begin{eqnarray}
\rho(x,\omega,t) = \sum_{l=1}^{m}
\rho_l^{(0)}(x -\Omega_l\omega_0 t,t) +
O(\omega_0^{-1}).
\label{h2}
\end{eqnarray}
The densities $\rho_l^{(0)}$ obey the following
Fokker-Planck equation
\begin{equation}
{\partial\rho_{l}^{(0)}\over\partial t} - T
{\partial^{2}\rho_{l}^{(0)}
\over\partial\beta^{2}} - K_0 \alpha_{l}
\,\frac{\partial}{\partial \beta}
\left\{ \rho_{l}^{(0)}\, \int_{-\pi}^{\pi}
f(\beta-\beta') \rho_{l}^{(0)}(\beta',t)
d\beta' \right\}  = 0,   \label{h3}
\end{equation}
where $\beta = x - \Omega_l\omega_0 t$ and each
$\int_{-\pi}^{\pi}\rho_{l}^{(0)} d\beta =1$. This
equation corresponds to the NLFPE (\ref{limit6})
- (\ref{limit7}) in the moving variable $\beta$,
with a coupling constant $K_0\alpha_l$ instead of
$K_0$ and $p(\omega) = \delta(\omega)$. We shall
see below that its solution evolves to a
stationary state as the time elapses provided
$f(x)$ is odd. Thus the probability density (to
leading order in $1/\omega_0$) is the sum of $m$
components obeying the stationary solution of
(\ref{h3}) with variables $x-\Omega_l\omega_0 t$.
The overall velocity (related to the order
function) is a superposition of $m$ waves each
traveling with frequency $\Omega_l \omega_0$
\begin{equation}
v(x,t) = - K_0\, \sum_{l=1}^{m} \alpha_l\,
\int_{-\pi}^{\pi} f(x-\Omega_l\omega_0 t-\beta')
\rho_{l}^{(0)}(\beta') d\beta' \,,
\label{h4}
\end{equation}
where $\rho_{l}^{(0)}(x)$ is the stationary
solution of (\ref{h3}).

We shall now consider NLFPE without  disorder,
i.e., $\omega_i=0$ or $p(\omega)=\delta(\omega)$.
Our results will be applicable to models with a
multimodal natural frequency distribution in the
high-frequency limit. For a model without
disorder, the moment-generating function is
independent of $y$ and it equals the  probability
density: $g(x,y,t) = \rho(x,0,t) \equiv g(x,t)$.
If detailed  balance is obeyed (i.e. if $f$ is an
odd function) then the model is purely
relaxational and consequently the formalism of
statistical mechanics can be  applied.


\section{Stationary solutions of models with odd
coupling}
\label{sec-odd}
We can find functional equations for the order
function of stationary or rotating wave
solutions at nonzero temperatures without making
Daido's assumptions. In this section we shall
assume detailed balance, which occurs when $f$
is an odd function extended periodically outside
$(-\pi,\pi)$ and satisfying $f(x+\pi) = -f(x)$.
(The expressions for the general case are
somewhat more  complicated). Let us define
\begin{eqnarray}
V(x,t) = \int_{-\pi}^x v(s,t)\, ds,\label{odd1}\\
W(x,\omega,t) = (\pi + x) \omega + V(x,t). \label{odd2}
\end{eqnarray}
The property $f(x+\pi) =-f(x)$ implies that the
drift (\ref{limit7})   satisfies $v(x+\pi,t) =-v(x,t)$, and
$V(x+\pi,t) = V(0,t) - V(x,t)$.


\subsection{Stationary solutions}
Let us restrict ourselves to the
case of stationary solutions;  rotating wave solutions may be
reduced to this case after moving to a   rotating frame. The
stationary solutions should have the form
\begin{eqnarray}
\rho(x,\omega) & = & Z^{-1}\, e^{{W(x,\omega)\over T}} \nonumber \\
& - &{J\over T} \int_{-\pi}^x
\exp\left[ {W(x,\omega) - W(s,\omega)\over T}\right]\, ds
,\label{odd3}
\end{eqnarray}
where $Z$ and $J$ are functions of $\omega$,
independent of $x$. We now impose   the condition that $\rho$ be a
$2\pi$-periodic function of $x$, use the symmetry properties of  the
drift, and find the probability flux $J$ as a function of $Z$:
\begin{eqnarray}
{J\over T} = {2\,\mbox{sinh}\left({2\pi\omega\over
T}\right)
\over Z \int_{-\pi}^{\pi} e^{-{W(x,\omega)\over T}}\, dx}\,.
\label{odd4}
\end{eqnarray}
$Z$ can be found from the normalization condition $\int_{-\pi}^{\pi}
\rho dx =1$. The functional equation for the drift $v(x)$ (or,
equivalently,  the order function) is obtained by inserting
(\ref{odd3}), (\ref{odd4})   and the formula for $Z$ in
(\ref{limit7}). The result is
\begin{eqnarray}
{v(x)\over K_{0}} = \int_{-\pi}^{\pi} f(x-x')
\int_{-\infty}^{\infty}    {e^{{W(x',\omega)\over T}}\over
Z(\omega)} \left\{ 2 \mbox{sinh}\left({2\pi \omega\over T} \right)
\right.\nonumber\\ \left.\times\, {\int_{-\pi}^{x'}
e^{-{W(s,\omega)\over T}}\, ds \over \int_{-\pi}^{\pi}
e^{-{W(s,\omega)\over T}}\, ds} - 1 \right\}\, p(\omega)\,
d\omega\, dx' , \label{odd41}
\end{eqnarray}
\begin{eqnarray}
Z(\omega) & = & \int_{-\pi}^{\pi}
e^{{W(x,\omega)\over T}}  \left\{   {2
\mbox{sinh}\left({2\pi\omega\over T} \right) \int_{-\pi}^{x}
e^{-{W(s,\omega)\over T}} ds\over\int_{-\pi}^{\pi}
e^{-{W(s,\omega)\over T}}\,   ds} - 1 \right\}\, dx .\nonumber
\end{eqnarray}
The functional equation (\ref{odd41}) for $v(x)$ may in general
have several   solutions, depending on $p(\omega)$ and the value
of the parameters $K_0$ and $T$. Notice that no assumption on the
shape of the order function has been   made in order to derive
(\ref{odd41}). Thus we have a general equation for $v(x)$ valid
for any temperature thereby extending Daido's theory. If $f(x)$ is
not odd, the same procedure yields a more cumbersome equation
which we omit.   An interesting case corresponds to the case
without disorder, $p(\omega) = \delta(\omega)$, for which $J=0$,
$\rho(x,0,t) = g(x,y,t) \equiv g(x,t)$, and
\begin{eqnarray}
g_0(x) = {e^{{V_{0}(x)\over T}}\over
\int_{-\pi}^{\pi}   e^{{V_{0}(s)\over T}}\, ds }\,. \label{odd5}
\end{eqnarray}
The subscript zero just reminds us that we are
considering stationary solutions. In this case
there is a general Liapunov functional related
to the free energy~\cite{Htheorem}. In Sections
\ref{sec-family}, \ref{sec-porous} and
\ref{sec-burgers}, we show that simpler quadratic
Liapunov functionals may exist for specific forms
of $f(x)$.


\subsection{Liapunov functional}
Let us define the relative entropy
\begin{eqnarray}
\eta(t) = \int_{-\pi}^{\pi} g(x,t)\,\ln\left({g(x,t)\over
\overline{g}(x,t)}
\right)\, dx ,\label{odd6}\\
\overline{g}(x,t) = e^{{V(x,t) - \mu(t)\over T}}\,,\label{odd7}\\
{d\mu\over dt} = \int_{-\pi}^{\pi} g(x,t)\, {\partial V(x,t)\over
\partial t}
\,  dx .\label{odd8}
\end{eqnarray}
$V(x,t)$ is given by (\ref{odd1}) with the drift calculated by means
of the   exact probability density $g(x,t)$. (\ref{odd8}) implies
that
$$\int_{-\pi}^{\pi} g(x,t)\, {\partial\over \partial
t}\ln\overline{g}(x,t)
\,  dx = 0.
$$
Direct calculation shows that
\begin{eqnarray}
{d\eta\over dt} = -T \int_{-\pi}^{\pi}
g(x,t)\,\left| {\partial\over\partial x}
\ln\left({g(x,t)\over \overline{g}(x,t)}\right) \right|^2 \, dx\leq
0  .\label{odd9}
\end{eqnarray}
The inequality $y\ln y\geq y-1$, $\forall y\geq 0$,
can be used to show that
\begin{eqnarray}
\eta \geq 1- \int_{-\pi}^{\pi} \overline{g}(x,t)\, dx .\label{odd10}
\end{eqnarray}
Then the relative entropy is bounded below if $
e^{V(x,t)} \in \mbox{L}^1(-\pi, \pi)
$  and the function $\mu(t)$ is  bounded below. The first condition
thanks to the Trudinger-Moser theorem (see \cite{BCS2}) is always
fulfilled if
$V(x,t)$
 is in the Sobolev space $W^{1,1}_0(-\pi, \pi)$ and this property is
verified for
$V(x, t)$ defined in (\ref{odd1}).  To show that the function
$\mu(t)$ is bounded below let us write Eq.\ (\ref{odd8}) as follows
\begin{eqnarray}
{d\over dt}\left[\mu - \int_{-\pi}^{\pi} g\, V\,
dx\right] =  -\int_{-\pi}^{\pi} V\, {\partial g\over \partial t}\,
dx \nonumber\\  = -\int_{-\pi}^{\pi} v\, \left[vg - T {\partial
g\over \partial x} \right]\, dx   = -\int_{-\pi}^{\pi} v(x,t)\,
J(x,t)\, dx ,
\label{cnu}
\end{eqnarray}
after using the nonlinear Fokker-Planck equation and
integration by parts.
$J$ $\equiv vg$ $- T\,\partial g/\partial x$ is the probability
flux. Then (\ref{cnu}) can be equivalently written as
\begin{eqnarray}
\mu(t) = \int_{-\pi}^{\pi} \left[ g(x,t)\, V(x,t)\, - g(x,0)\,
V(x,0)\right] dx \nonumber \\
-  \int_0^t \int_{-\pi}^{\pi} v(x,s)\,
J(x,s)\, dx \ ds.
\label{cnu1}
\end{eqnarray}
Let us first note that, since $\int_{-\pi}^{\pi} g \, dx = 1$ and
$V$ is a uniformly bounded function, the two first terms in the
right hand side of (\ref{cnu1}) are bounded. If we can prove that
$$
\int_0^t \int_{-\pi}^{\pi} v(x,s)\,
J(x,s)\, dx \ ds
$$
is uniformly bounded, then
$\mu(t)$ will remain bounded for  all time and the demonstration
that $\eta$ is a Liapunov functional will be  over. By using the
symmetry of $f(x)$ and the definitions of $J$ and $v$, it is easy
to show first that
$$
\int_{-\pi}^{\pi} J(x,t)\, dx = 0, \quad \forall t.
$$
Since $J$ is a continuous function with respect to $x$, the mean
value theorem applied to the above equality implies the existence
of $a_t \in [-\pi,\pi]$ such that $J(a_t,t)=0$, where the subscript
makes reference to the fixed time $t$ and $a_t$ is chosen so that
the function $J$ has a definite sign over $(a_t,x)$. Then
$$
\int_0^t \int_{a_{t}}^{x} v(x,s)\,
J(x,s)\, dx \ ds \leq \mbox{max}_{x,t} |v(x,t)|\,\int_{a_{0}}^{x}
\left[\mbox{sign} J(x,t)\,\int_0^{b_{x}} J(x,s)
ds\right]\, dx,
$$
where $t=b_x$ is the inverse function of $x=a_t$.
Now we integrate
$$
{\partial g \over \partial t} = -T {\partial J \over \partial x}
$$
over $(a_0, x) \times (0,b_x)$, and obtain
$$
\int_{a_t}^x [g(y,t) - g(y,0)] \, dy = -T\,\int_0^t\int_{a_t}^x
{\partial J(y,s)\over\partial y}\, dy ds = -T\,\int_0^t J(x,s) ds,
$$
where we have used again Fubini's theorem to exchange the order of
integration in the right hand side. The left hand side of this
equation is bounded by 2, so that we obtain after taking the limit
as $t\to\infty$,
$$
\lim_{t \to \infty} \int_0^t J(x, s) \, ds \ \leq \ {1 \over
T} \lim_{t
\to
\infty} \int_{a_t}^x [g(y,t) - g(y,0)] \, dy \ \leq \ {2 \over T}.
$$
Combining this bound with the uniform one for $v(x,t)$ and taking
into account that
$$
\int_0^t \int_{-\pi}^{\pi} v(x,s)\,
J(x,s)\, dx \, ds \, = \, \sum_{j \in \Lambda} \int_0^t
\int_{a_{t}^j}^{x} v(x,s)\, J(x,s)\, dx \, ds,
$$
where $\Lambda$ is a finite set of indices because of $J\in C^1$ has
a finite number of zeros in $[-\pi, \pi]$, we deduce that the third
term in (\ref{cnu1}) is also uniformly bounded and, as consequence,
$\mu(t)$ is bounded from below.

Our Liapunov functional may be used to show that $g(x,t)$ tends to a
stationary   solution as time elapses and also to discuss the global
stability of these   solutions. Let $g_\infty(x,t)$ be the limiting
probability density as $t\to
\infty$. Equating the entropy production (\ref{odd9}) to zero, we
obtain
$g_\infty = \beta(t)$ exp$V(x,t)/T$. Inserting this into
(\ref{limit6}), we find
$$
{d\beta\over dt} + {\beta\over T}\, {\partial V(x,t)\over \partial
t} = 0,
$$
which may be rewritten as
$$
{T\over \beta} {d\beta\over dt} = - {\partial V(x,t)\over \partial
t}\,.
$$
The left side of this expression is independent of $x$
whereas the second side  is not, so that both sides are zero. Then
$V$ is time independent and $\beta$  is a constant, which proves
$g_\infty$ to be of the form (\ref{odd5}).


\subsection{Equilibrium states}
The previous considerations suggest that for models with
odd-coupling functions and no frequencies [i.e. $f(-x)=-f(x)$ and
$p(\omega)=\delta(\omega)$], a thermodynamic formulation will
suffice to identify the stationary states as well as the possible
existence of thermodynamic singularities (i.e., bifurcations).

In order to demonstrate this assertion, we shall start by defining
an appropriate energy function,
\begin{eqnarray}
{\cal H}=\frac{K_0}{N}\,\sum_{i<j}
E(\phi_i-\phi_j)
\label{f1}
\end{eqnarray}
where $E(x)$ is a two-pair interaction energy
function defined by $E(x)=\int_{-\pi}^x f(s) ds$ and
$E(x)=E(x+2\pi)$. The $2\pi$-periodicity of $E(x)$ is crucial to derive the
results of this section.  Note that this newly introduced function is
related to the potential $V(x,t)$ in Eq.\ (\ref{odd1}) by
\begin{eqnarray}
 - {V(x,t)\over K_{0}} =
\int_{-\infty}^{\infty}\int_{-\pi}^{\pi} E(x-x') \rho(x',\omega,t)
p(\omega) dx' d\omega \nonumber\\  -
\int_{-\infty}^{\infty}\int_{-\pi}^{\pi} E(-\pi-x')
\rho(x',\omega,t)   p(\omega) dx' d\omega . \label{f2}
\end{eqnarray}
which has the physical meaning of an averaged
energy. The second term is an   additive constant which fixes the
origin of the energy scale. Consequently the  Liapunov functional
defined in Eq.\ (\ref{odd6}) is a generalized free energy  for the
system. Now we want to show that the equilibrium state of such a
system yields potential solutions identical to (\ref{odd5}). Let us
compute the partition function,
\begin{eqnarray}
 {\cal Z} = \int_{-\pi}^{\pi}\ldots
\int_{-\pi}^{\pi} e^{-\beta{\cal H}}\,
\prod_{i=1}^N d\phi_i \quad\quad\quad\quad\quad\quad\quad\quad\quad
\nonumber\\  = \int_{-\pi}^{\pi}\ldots \int_{-\pi}^{\pi}
\exp\left(-\frac{\beta
K_0}{N}\sum_{i<j}E(\phi_i-\phi_j)\right)\,\prod_{l=1}^N d\phi_l .
\label{f3}
\end{eqnarray}

We have $a_0 = 0$ because $f$ is odd. Then we can write $E(x) =
\sum_{n=-\infty}^{\infty} b_n e^{inx}$ (plus an unessential
constant term),  where $b_0=0$, $b_n=-i(a_n/n)\,(n\neq 0)$.
Consequently we can rewrite  the partition function for an
oscillator system without frequency disorder,
$p(\omega)=\delta(\omega)$, as
\begin{eqnarray}
{\cal Z}=\int_{-\pi}^{\pi}\ldots
\int_{-\pi}^{\pi}
\exp\left\{ -\frac{\beta K_0}{2N} \sum_{n=-\infty}^{\infty}
\left(\sum_{i=1}^N e^{in\phi_{i}}
\sum_{j=1}^N e^{-in\phi_{j}} \right) \right\}\,
\prod_{l=1}^N d\phi_l .
\label{f4}
\end{eqnarray}
 We have neglected the contribution of the terms
$(i=j)$ in the exponential  (which is equivalent to redefine the
origin of energies so that $E(0)=0$).   Let us now insert delta
functions in the  previous integrals and use the identity,
\begin{eqnarray}
1 &=& \int_{-\infty}^{\infty}
\delta\left(h_n-\frac{1}{N}\sum_{i=1}^N   e^{in\phi_{i}}\right) \,
dh_n \nonumber\\  &=& \int_{-\infty}^{\infty}\int_{-\infty}^{\infty}
\exp\left\{i\lambda_n\left(h_n-\frac{1}{N}\sum_{i=1}^N
e^{in\phi_{i}}
\right)\right\}\, dh_n d\lambda_n ,\nonumber
\end{eqnarray}
where the new set of Lagrange multipliers
$\lambda_n$ has been  introduced. Inserting these representations of
the delta function in  (\ref{f4}) and permuting the orders of the
integration  between the $\phi's$ and the $h_n,\lambda_n$, we reduce
the final expression  to a single site problem,
\begin{eqnarray}
{\cal Z}=\int_{-\infty}^{\infty}
\exp [N A(\lambda,h)]\,\prod_{n=-\infty}^{\infty} d\lambda_n dh_n ,
\label{f6}
\end{eqnarray}
where
\begin{eqnarray}
A(\lambda,h)=-\sum_{n=-\infty}^{\infty}\lambda_n
h_n -\frac{\beta K_0}{2}
\sum_{n=-\infty}^{\infty} b_n h_n h_{-n} \nonumber\\  + \log
\int_{-\pi}^{\pi}
\exp\left(\sum_{n=-\infty}^{\infty}\lambda_n e^{in\phi} \right)\,
d\phi .
\label{f7}
\end{eqnarray}

The dominant contribution to the partition function is determined
by the saddle point equations,
\begin{eqnarray}
\frac{\partial A}{\partial \lambda_n}=\frac{\partial A}{\partial
h_n}=0
\label{f8}
\end{eqnarray}
which yield,
\begin{eqnarray}
h_n &=& <\exp(in\phi)> , \label{f9a}\\
\lambda_n  &=& -\frac{\beta K_0}{2} h_{-n}(b_n+b_{-n}) = -\beta K_0
h_{-n}
\,\mbox{Re}(b_n)\,.
\label{f9b}
\end{eqnarray}
Here the average $<B(\phi)>$ of an arbitrary
function
$B(\phi)$ is defined as follows
\begin{eqnarray}
<B(\phi)>=\frac{\int_{-\pi}^{\pi} B(\phi)\, \exp
\bigl(\sum_{n=-\infty}^{\infty}\lambda_n e^{in\phi}\bigr)\, d\phi}
{\int_{-\pi}^{\pi}\exp\bigl ( \sum_{n=-\infty}^{\infty}
\lambda_n e^{in\phi}\bigr )\, d\phi}.
\label{f10}
\end{eqnarray}

The next steps are quite standard. Inserting the expression
(\ref{f9b})   for $\lambda_n$ into (\ref{f9a}), we obtain a closed
set of equations for  the parameters $h_n$. The reader will easily
convince himself that the  parameters $h_n$ are nothing less that
the moments defined in  Eq.\ (\ref{eq3}) (with $\omega_i=0$).
Finally we find that the effective  Hamiltonian in the exponent of
(\ref{f10}) is exactly given by the potential  function $V_0(x)$
of eq.(\ref{odd1}). Then the equilibrium density function $g(x)$
should satisfy
\begin{eqnarray}
g(x)=<\delta(x-\phi)>=\frac{\exp(\beta
\tilde{V}(x))}  {\int_{-\pi}^{\pi} \exp(\beta \tilde{V}(x))\, dx}\,,
\label{f11}
\end{eqnarray}
with
\begin{eqnarray}
\tilde{V}(x)=-K_0\sum_{n=-\infty}^{\infty} (Re b_n)h_{-n}\exp(inx)
dx.
\label{12}
\end{eqnarray}
This coincides with the definition (\ref{odd1})
[$\tilde{V}(x)=V(x)+$constant]   if all the $b_n$ are real i.e. if
$f(x)$ is an odd function. From (\ref{f9a})   and (\ref{f9b}) it
follows that the function $v(x)$ is a solution of the  following
functional equation
\begin{eqnarray}
 v(x) = - K_0 <f(x-\phi)>  =
-K_0\frac{\int_{-\pi}^{\pi}  f(x-\phi)\, e^{\beta V(\phi)}
d\phi}{\int_{-\pi}^{\pi}   e^{\beta V(\phi)}\, d\phi} .
\label{f13}
\end{eqnarray}
The system of equations (\ref{odd1}) and (\ref{f13})
may have one, many or   no solutions. In case of multiple solutions,
the free energy and the   dynamics should be used to fully
characterize the equilibrium  state.



\section{A family of models}

\label{sec-family}

When $p(\omega)=\delta(\omega)$, Eq.(\ref{eq6})
becomes
\begin{equation}
\frac{\partial g}{\partial t} = -\frac{\partial}{\partial x}
\Bigl [v(x,t)g \Bigr ]+T\frac{\partial^2 g}{\partial x^2}
\,.\label{eq8}
\end{equation}
The probability density $g(x,t)$ obeys the natural
normalization condition
\begin{equation}
\int_{-\pi}^{\pi} g(x,t) \, dx = 1\,.\label{eq.norm}
\end{equation}

Let us consider the following {\it special} cases of localized
coupling functions $f$:
\begin{itemize}
\item (a) {\em The hard-needles model}: In this case $f = \delta'$
and (\ref{eq6}) and (\ref{conv}) yield
\begin{equation}
\frac{\partial g}{\partial t} = K_0\frac{\partial}{\partial x}
\Bigl (g\frac{\partial g}{\partial x}\Bigr ) + T \frac{\partial^2
g}{\partial x^2}\,.\label{eq9}
\end{equation}
 Note that in this case there is an energy of the
model given by $E=\sum_{i<j} \delta(\phi_i-
\phi_j)$ which is trivially zero except when two
phases coincide. If we imagine the phases as
needles located in the centre of the unit circle
then the crossing of needles costs infinite
energy. This is the reason for its name: In this
case the needles are impenetrable and interact
like the hard spheres in the theory of liquids.
We shall see later that synchronization for this
model is not possible at any temperature $T\geq
0$: any initial configuration $g(x,0)$ evolves
towards incoherence [$g=(2\pi)^{-1}$ which means
that all angles have the same probability to
occur] for large enough time. This result is
anticipated by considerations of linear stability
alone. In fact, linearizing  (\ref{eq9}) about
$(2\pi)^{-1}$ we obtain the equation $$
{d^{2}\hat{g}\over dx^{2}} - {2\pi\lambda\over
K_{0} + 2\pi T}\,\hat{g} = 0, $$ for $g(x,t) =
(2\pi)^{-1} + \hat{g}(x)\, e^{\lambda t}$. This
equation has $2\pi$-periodic solutions of zero
mean if $-\lambda = T + K_0/(2\pi)$. Thus
incoherence is linearly stable for any $T\geq 0$.
In addition, as we will see in the next section,
there exists a Liapunov functional which shows
that $g(x,t)$ should evolve in
$L^1(-\pi, \pi)$ towards $x$ independent functions which, by
normalization, is equal to $(2\pi)^{-1}$.
\item (b) {\em The stick-needles model}: Now $f=- \delta'$ and we
obtain the following evolution equation for $g$:
\begin{equation}
\frac{\partial g}{\partial t} = - K_0\frac{\partial}{\partial x}
\Bigl (g\frac{\partial g}{\partial x}\Bigr ) + T \frac{\partial^2
g}{\partial x^2}\,.\label{eq10}
\end{equation}
The only difference with the previous case is the
change of sign in the velocity field. This is enough to
dramatically alter the behaviour of the model. Now the energy is
given by $E=-\sum_{i<j} \delta(\phi_i-\phi_j)$. Obviously the
thermodynamics is badly defined because the ground state energy is
$-\infty$. The needles want to stick to each other and there is
only one relevant configuration which dominates the partition sum.
Linear stability considerations indicate that incoherence is
stable only if $T>K_0/(2\pi)$. In general this model is not
mathematically well-posed unless $T-K_0 g(x,t) \geq 0$.
\item (c)  {\em The Burgers model}: If $f= \delta$, the velocity
becomes $v(x,t)=-K_0 g$ and the dynamical equation for $g$ is
\begin{equation}
\frac{\partial g}{\partial t} = 2 K_0 g\frac{\partial g}{\partial x}
+ T \frac{\partial^2 g}{\partial x^2}\,.\label{eq11}
\end{equation} The model corresponds to the Burgers equation (BE).
Note that the two coupling functions $f=\pm \delta$ are both
equivalent (to go from the + to the - case or viceversa it is enough
to make the transformation $x\to -x$.).  The dynamics of the model
corresponds to the Burger equation with a supplementary periodic
boundary condition for the $g$ (which plays the role of the velocity
field in the BE equation). Performing the transformation $x\to
x'=-x/2K_0$ we obtain the Burgers equation
\begin{equation}
\frac{\partial g}{\partial t} + g\frac{\partial g}{\partial x}
=\nu\frac{\partial^2 g}{\partial x^2},
\label{eq12}
\end{equation}
with viscosity $\nu=\frac{T}{4K_0^2}$. Physically
this model corresponds to a system of needles
which tend to move together in the same direction
when they meet. The incoherent solution is always
a stationary solution of the BE equation.
Straightforward but tedious calculation shows
that {\em the incoherent solution is the only
stationary solution satisfying periodicity and
normalization conditions}. Notice that the coupling
function for the Burgers model is not odd
and therefore the general functional (\ref{odd6})
of Section 3 may no longer be a Liapunov
functional. However, the functional
$G(t)={1\over 2}\int_{-\pi}^{\pi} g(x,t)^2 dx$
satisfies $G'(t) = -T \int_{-\pi}^{\pi} (\partial
g(x,t)/\partial x)^2 dx$, and is therefore a
Liapunov functional for $T>0$, but not at zero
temperature. Thus any initial configuration
evolves towards incoherence for $T>0$, while in
principle synchronization is possible only at
zero temperature. These results agree with
considerations of linear stability for the
incoherent solution: the  probability density
$g= 1/(2\pi)$ is linearly stable for $T>0$ and
neutrally  stable for $T=0$. Finally, we shall
mention that this model, contrarily to what
happens in models  (a) and (b), lacks a
thermodynamic formulation because its dynamics
violates  detailed balance.
\item (d)  {\em Daido coupling (the extended Burgers model)}: If we
choose
$f(x) \ = \
\mbox{sign}(x)$ (periodically extended outside the interval
$[-\pi,\pi]$) as coupling function, the expression (\ref{conv}) for
the velocity becomes
\begin{eqnarray}
v(x,t) \ = \ - K_0 \,\int_{-\pi}^{\pi}\,
\mbox{sign}(\xi) \ g(x-\xi,t) \ d\xi
 \ \nonumber\\ = K_0 \left[\int_{-\pi}^{0} g(x-\xi,t) d\xi -
\int_{0}^{\pi} g(x-\xi,t) d\xi \right] \nonumber\\ = -  K_0
\int_{0}^{\pi} \left[ g(x-\xi,t) - g(x- \xi+\pi,t)
\right] \ d\xi. \label{v(x,t)}
\end{eqnarray}
Then we have
\begin{eqnarray}
{\partial v \over \partial x} = -2K_0 \, [g(x,t) -
g(x+\pi,t)].
\label{dV}
\end{eqnarray}
The moment-generating function satisfies (\ref{eq8})
and (\ref{v(x,t)}) which form a nonlocal equation for $g(x,t)$.
Synchronization already appears at non-zero temperature ($T > 0$)
for this simple model which is purely relaxational (when all the
oscillator frequencies are equal to zero). The corresponding
Hamiltonian is given by
\begin{equation}
{\cal H}=\frac{K_0}{N}\sum_{i<j} mod(|\phi_i-\phi_j|,2\pi) .
\label{eq13}
\end{equation}
\end{itemize}
In what follows we are going to analyze the dynamical behaviour
of his family of models.


\section{ The porous Medium models}

\label{sec-porous}

\subsection{The hard-needles model}
The dynamical equation for the
hard-needles model is
\begin{equation}
{\partial g \over \partial t} \  = \  K_0{\partial
\over \partial x} \left(g {\partial g \over \partial x} \right) \ +
\ T{\partial^2 g \over \partial x^2}
\label{PM+}
\end{equation}
which is considered together with periodic boundary
conditions
\begin{equation}
g(-\pi,t) = g(\pi,t) \ \mbox{ and } \ {\partial g
\over \partial x}(-\pi,t)  = {\partial g \over \partial x}(\pi,t)
\label{bc}
\end{equation}
and the initial condition
\begin{equation} g(x,0) \ = \ g_0(x).
\label{ic}
\end{equation}


\subsubsection{Well-possed problem.}

This equation is a particular case of a the general quasi-linear
equation in divergence form studied by O. A. Lady\v zenskaja, V. A.
Solonnikov and N. N. Ural'ceva in \cite{LSU}, Theorem 6.1 and by S.
N. Kruzhkov in \cite{K1}. As a consequence of the results in
\cite{LSU} and \cite{K1} we can deduce that the problem is
well-posed if
$$
 K_0 g \ + \ T \geq 0,
$$ which is always fulfilled for $g \geq 0$.


\subsubsection{Decay in time estimates.}

In order to obtain some decay estimates in time for the solutions of
solution (\ref{PM+}) towards its equilibrium state, we must obtain a
Green function to the linear heat equation
\begin{equation}
{\partial g \over \partial t} \  = \  T{\partial^2
g \over \partial x^2}
\label{lh}
\end{equation}
with periodic boundary conditions (\ref{bc}). Let
$\Gamma(x,t)$ be the heat kernel, i.e., the fundamental solution of
the heat equation in $\RR$, defined by
$$
\Gamma(x,t) \ = \ (4\pi Tt)^{-{1 \over 2}} e^{- {|x|^2 \over 4Tt}} .
$$
Let us recall the definition of the  Theta function \cite{Fo},
which is related to one of Jacobi's elliptic functions~\cite{AS},
\begin{equation}
\Theta(x,t) \ = \ \sum_{n=-\infty}^{\infty} \Gamma(x + 2n\pi,t) ,
\label{theta}
\end{equation}
for $t > 0$, and $\Theta = 0$ for $t < 0$, and define
the function $\Psi$ by
\begin{equation}
\Psi(x,t) \ = \ \sum_{n=-\infty}^{\infty} {\partial \Gamma \over
\partial x} (x + 2n\pi,t).
\label{psi}
\end{equation}
The functions $\Theta$ and $\Psi$ satisfy
\begin{equation}
\Theta(x + 2\pi,t) = \Theta(x,t)  \, \mbox{ and } \, \Psi(x +
2\pi,t) =
\Psi(x,t).
\label{thetabc}
\end{equation}

Then, the function $\Theta(x-z,t-s)$ verifies: (i) it is a solution
of the  linear heat equation (\ref{lh}) except in $(s,z)$; (ii) it
satisfies (\ref{bc});
$\Theta(x-z,t-s) - \Gamma(x-z,t-s)$ is solution of (\ref{lh}). As a
consequence,
$\Theta(x-z,t-s)$ is a Green function for the linear heat equation
with periodic  boundary conditions. Another interesting property of
the $\Theta$ function is  its positivity. Also, note that the
$\Theta$ function can be written as follows
$$
\Theta(x,t) = {1 \over 2\pi} + {1 \over \pi}\sum_{n=1}^{\infty}
e^{-n^2 T t} \cos nx .
$$

In view of the preceding properties, we can write the solution of
the  hard-needles model in the following integral form
\begin{eqnarray}
 g(x,t) = \int_{-\pi}^{\pi}\Theta(x-z,t)  g_0(z)\
dz\quad\quad\quad\quad\quad\quad
\quad\quad \nonumber\\ +  \int_{-\pi}^{\pi}\int_0^t {\partial
\Theta(x-z,t-s) \over \partial z}
\left( g {\partial g \over \partial z} \right) (z,s) \ dz \ ds.
\label{i1}
\end{eqnarray}

We will  use   for $-\gamma > -{1 \over 2}$, $t>0$  and
$x>0$ the following estimate (see \cite{W})
\begin{equation}
\left| {\partial^k \Gamma(x,t) \over \partial x^k}\right| \ \leq \
{c\,  (Tt)^{- \gamma} \over |x|^{1-2\gamma  + k}},\quad k = 0,1,2,
\ldots ,
\label{bgamma}
\end{equation}
where $c= (4\pi)^{-1/2}(-4\gamma e^{-1} + 2
e^{-1})^{(1-2\gamma)/2}$ and we have chosen $-\gamma$ such that
$-1/2 < - \gamma < 0$ and close enough to
$-1/2$.

Using these estimates, the generating function and its derivative
with respect to the space variable can be estimated from (\ref{i1})
for small initial data
$g_0
\in L^{1}_{2\pi}$, where $L^{p}_{2\pi}$ denotes the space of
$2\pi$-periodic functions belonging to $L^p(-\pi,\pi)$, see
\cite{Ve} for definition and main properties. This procedure was
introduced in
\cite{CS} by G. H. Cottet and J. Soler to study the decay properties
of the Navier-Stokes equations with weak initial data (singular
filament measure). Let $t \in [0,\tau]$, $\tau \in\RR$. Firstly, due
to the positivity of
$\Theta$, $g_0$ and  of the solution, from equation (\ref{PM+}) we
deduce
\begin{equation}
\| g(\cdot,t) \|_{L^1_{2\pi}} \ = \ \int_{-\pi}^{\pi} g_0(x) \ dx.
\label{l1}
\end{equation}

To obtain the $L^p_{2\pi}$ estimates of the solution let us note
that since
$\Theta$ converges uniformly we have
\begin{eqnarray}
\int_{-\pi}^{\pi}\Theta(x-z,t)  g_0(z) \ dz \ - \ {1 \over 2\pi}
\int_{-\pi}^{\pi} g_0(z) \ dz \nonumber\\ = \ {1 \over
\pi}\sum_{n=1}^{\infty} e^{-n^2 T t} \int_{-\pi}^{\pi} \cos n(x-z)
\; g_0(z) \ dz. \nonumber
\end{eqnarray} In the case of the uniform bound for $g(x,t)$, we
estimate (\ref{i1}) as follows
\begin{eqnarray}
\left\| g(\cdot,t) - {\int_{-\pi}^{\pi} g_0(x) dx\over 2\pi}
\right\|_{L^\infty_{2\pi}} \leq \left\|\Theta(\cdot,t) - {1 \over
2\pi}
\right\|_{L^\infty_{2\pi}}
\| g_0 \|_{L^1_{2\pi}} \nonumber\\ + \int_0^t  \left\| {\partial
\Theta \over \partial x}(\cdot,t-s)\right\|_{L^{p'}_{2\pi}} \|
g(\cdot,s) \|_{L^\infty_{2\pi}}
\left\| {\partial g \over \partial x}(\cdot,s)
\right\|_{L^p_{2\pi}} ds .
\label{Linf}
\end{eqnarray}
Also, we have
\begin{eqnarray}
\left\| g(\cdot,t) - {\int_{-\pi}^{\pi} g_0(x) dx\over 2\pi}
\right\|_{L^p_{2\pi}} \leq \left\|
\Theta(\cdot,t) - {1 \over 2\pi} \right\|_{L^p_{2\pi}}
\| g_0 \|_{L^1_{2\pi}} \nonumber\\ + \ \int_0^t  \| {\partial \Theta
\over \partial x}(\cdot,t-s)
\|_{L^{1}_{2\pi}}\| g(\cdot,s) \|_{L^\infty_{2\pi}} \| {\partial g
\over
\partial x}(\cdot,s) \|_{L^p_{2\pi}} ds,
\label{Lp}
\end{eqnarray}
and
\begin{eqnarray}
\left\| {\partial g \over \partial x}(\cdot,t) \right\|_{L^p_{2\pi}}
\leq
\left\| {\partial \Theta \over  \partial x}(\cdot,t)
\right\|_{L^p_{2\pi}}
\|  g_0  \|_{L^1_{2\pi}} \  + \ \int_0^t  \left\|
{\partial\Theta\over\partial x}(\cdot,t-s)
\right\|_{L^{p'}_{2\pi}}
\nonumber\\
\times \left( \left\| {\partial g \over \partial x}(\cdot,s)
\right\|_{L^{p}_{2\pi}}^2  +
\| g(s, \cdot) \|_{L^\infty_{2\pi}} \left\| {\partial^2 g\over
\partial x^2}(\cdot,s) \right\|_{L^{p /2}_{2\pi}} \right)  ds.
\label{DLp}
\end{eqnarray}
To close the circle of these estimates requires to
prove that ${\partial^2 g
\over
\partial x^2} \in L^{p/2}_{2\pi}$. This can be obtained directly
from the equation (\ref{PM+}) combined with an estimate of
${\partial g
\over \partial t} \in L^{p/2}_{2\pi}$ deduced also from (\ref{i1}).
We bound  the previous inequalities using (\ref{bgamma}).

Let $M_p(t)$ be, with $1 < p \leq \infty$, such that
\begin{eqnarray}
{(Tt)^{\gamma}\over 1 + C\, (Tt)^{1-2\gamma}}
\left\| g(\cdot,t)
 - {1\over 2\pi}\int_{-\pi}^{\pi} g_0(x) dx \right\|_{L^p_{2\pi}}
\leq M_p(t), \nonumber\\ {(Tt)^{\gamma}\over 1 + C\,
(Tt)^{1-2\gamma}} \left\| {\partial g
\over \partial x} (\cdot,t) \right\|_{L^p_{2\pi}} \leq
M_p(t).\nonumber
\end{eqnarray}
Thus $\| {\partial g\over \partial x} (\cdot,t)
\|_{L^p_{2\pi}}$ tends to 0 as $t\to +\infty$ if we choose $- \gamma
< -1/3$.

Set $M = \mbox{max} \{ M_p(t), \ 1 \leq p \leq \infty, \ t \in [0,
\tau] \}$. Then, we have for (\ref{Linf}), (\ref{Lp}) and
(\ref{DLp}) the quadratic equation
\begin{equation}
M \leq C(g_0) + C(\tau)M^2 ,
\label{M}
\end{equation}
where $C(g_0)$ is a constant depending on the norm of
$g_0$ in $L_{2\pi}^{1}$.
 Inequality (\ref{M}) also implies at least local existence for any
initial data or global existence of the solution for small initial
data in $L_{2\pi}^{1}$. For small initial data in $L_{2\pi}^{1}$ and
$t \in [0, \tau], \
\forall \tau \in \RR$, (\ref{M}) yields
\begin{eqnarray}
\left\| g(\cdot,t)   - {1\over 2\pi}\int_{-\pi}^{\pi} g_0(x)\, dx
\right\|_{L^p_{2\pi}}
\leq M (Tt)^{-\gamma}\left[ 1 + C\, (Tt)^{1-2\gamma} \right],
\label{ep}\\
\left\| {\partial g \over \partial x} (\cdot,t)
\right\|_{L^p_{2\pi}} \leq M (Tt)^{-\gamma}\left( 1 + C\,
(Tt)^{1-2\gamma} \right).
\label{edp}
\end{eqnarray}
There exists other methods to study the existence
properties but they do not give the above sharp estimates
(\ref{ep})-(\ref{edp}), see \cite{LSU} and \cite{K1}.


\subsubsection{Asymptotic behaviour as $t \to \infty$.}

Let us obtain that the system is simplified asymptotically as $t\to
\infty$. In fact, we will prove that the solution of equation
(\ref{PM+}) converges in
$L^1_{2\pi}$ as $t \to \infty$ towards a limit function by using a
Liapunov functional associated to the system. The  functional
\begin{equation}
G(t) = {1\over 2}\,\int_{-\pi}^{\pi} g(x,t)^2\,
dx,\label{lia1}
\end{equation}
is positive and it satisfies
\begin{equation}
G'(t) = - \int_{-\pi}^{\pi} (T + K_0 g)\,
\left({\partial g\over\partial x}
\right)^2 dx \leq 0.\label{lia2}
\end{equation}
Thus it is a Liapunov functional and $g(x,t)$ should
evolve towards $x$ independent functions which, by normalization,
equal $(2\pi)^{-1}$. Since we have developed in the general case the
study of the properties of the Liapunov functional, we refer to
Section \ref{sec-odd} for the detailed arguments of the above
assertions.


\subsubsection{Convergence towards the solution of the Porous Medium
Equation  as $T \to 0$.}

 We want to solve equation (\ref{PM+}) from another point of view,
and study the behaviour of its solutions as $T \to 0$: we propose
an argument of splitting in time (with step $\Delta t$) between
the heat equation and the Porous Medium Equation (PME). Let us
first consider the periodic porous Medium problem
\begin{equation}
{\partial g^{PM} \over \partial t} \ = \ K_0
{\partial \over \partial x}
\left(g^{PM} {\partial g^{PM} \over \partial x} \right) ,
\label{PME}
\end{equation}
\begin{eqnarray}
g^{PM}(-\pi,t) = g^{PM}(\pi,t) \quad\quad
\mbox{and} \quad\quad {\partial g^{PM} \over
\partial x}(-\pi,t) = {\partial g^{PM} \over \partial x}(\pi,t),
\label{PMEcc}\\ g^{PM}(x,0) \ = \ g_0(x).
\label{PMEci}
\end{eqnarray}

The initial and initial-boundary value problems
associated  to this equation  were first studied
around 1950 by Zel'dovich, Kompaneets and
Barenblatt (see \cite{ZK} and \cite{B}). These
authors explicitly obtained a fundamental
(self-similar) solution of (\ref{PME}) in $\RR$:
\begin{equation}
Z\left({x\over K_{0}},t\right) = t^{-{1\over 3}}
\left[ C - {|x|^{2}\over 12 t^{{2\over 3}}} \right]_{+}
\,,\label{self-similar}
\end{equation}
where $[s]_+ =$ max$\{ s,0\}$ and $C$ is an
arbitrary positive constant. This solution has
compact spatial support for every fixed positive
time. There is a free-boundary or propagation
front, given by
$$
t = (12 C)^{-{3\over 2}}\, |x|^3\,,
$$
which separates regions where $g^{PM}>0$ from
those where $g^{PM}=0$. (\ref{self-similar})
presents corners (jump discontinuities in its
first derivatives) on the free boundary. In the
1-D case, the study of these problems for the
PME  started with the results of Oleinik and
coworkers in \cite{OKC}. See  the surveys by A. S.
Kalashnikov \cite{Ka} or by J. L. V\'azquez
\cite{V} for more information. However, to our
knowledge, the periodic boundary problem  has not
been studied so far.

It is known, see \cite{V,P}, that for all
initial data $g_0 \in L^1_{2\pi}$ there exists a
unique weak solution $g^{PM} \in C([0,\infty);
L^1_{2\pi})$, which is not classical in general.
This is illustrated nicely by the self-similar
solution (\ref{self-similar}), which is
$2\pi$-periodic. In our periodic boundary context,
this selfsimilar solution makes sense, at least
locally for $0<t<\pi^3/(12 C)^{{3\over 2}}$,
before the propagation front arrives to the
boundary. This example illustrates that a
solution (corresponding to arbitrary initial data)
may have jump discontinuities in its first
derivatives on the  propagation front or free
boundary separating the regions where the
solution  is positive from those where
$g^{PM}=0$. Moreover, disturbances propagate
with  finite speed. In general, it is clear that
$g^{PM}\geq 0$ for $t>0$, provided $g_0\geq 0$.
Solutions of (\ref{PME})-(\ref{PMEci}) have the
following behaviour as $t \to \infty$:
\begin{eqnarray}
\lim_{t \to \infty} \left\|g^{PM}(\cdot,t) - { \int_{-\pi}^{\pi}
g_{0}(x)  dx\over  2\pi} \right\|_{L^p(-\pi,\pi)}
= 0, \quad \forall p \in [1, \infty].
\label{abpme}
\end{eqnarray}
This result can be proved, according to C.\ M.\
Dafermos's ideas \cite{Da}, by following the basic steps listed
below:
\begin{itemize}
\item[1)] Prove that the semigroup $U$ corresponding to the problem
(\ref{PME})-(\ref{PMEci}) is a continuous contraction semigroup on
$L^1(-\pi,\pi)$ which verifies the maximum principle
$$
\| U(t)g_0 \|_{L^p(-\pi,\pi)} \ \leq  \  \| g_0 \|_{L^p(-\pi,\pi)}
$$
and
$$
g_0 \leq \bar{g}_0 \ \mbox{a.e.\ in}\, [-\pi,\pi], \
\mbox{implies} \ U(t)g_0
\leq U(t)\bar{g}_0 .
$$
\item[2)] The orbit $\gamma(g_0) = \cup_{t\geq 0}U(t)g_0$ is
relatively compact in $L^1(-\pi,\pi)$ for $g_0 \in
L^\infty(-\pi,\pi)$.
\item[3)] The $\omega$-limit set $\omega(u_0)$ is non-empty and
compact in
$L^1$.
\item[4)] Use an appropriate Liapunov functional such as
$$
V(\xi) \ = \ \mbox{ess sup}_{x \in [-\pi,\pi]} \xi(x), \ \ \xi
\in L^1(-\pi,\pi), $$ and the contractive property of the
associated semigroup to prove that (i) $V$ is a constant $W$ on
$\omega(u_0)$, and (ii) the $\omega$-limit set consists of
constants.
\item[5)] The comparison principle given in 1) and the fact that the
average of the solution is preserved allows us to identify $W$ with
$$
{1 \over 2\pi} \int_{-\pi}^{\pi} g_0(x) \ dx.
$$
and to prove (\ref{abpme}) for $p=1$.
\end{itemize}
The result follows for $p >1$ by using the dominated
convergence theorem.

A similar proof based on Dafermos's theory was introduced by N. D.
Alikakos and R. Rostamian in \cite{AR} in the case of boundary
condition of type
$$ {\partial (g^{PM})^2 \over \partial x}(x) \ = \ 0, \ \mbox{ for }
\ x \ = -\pi,
\,\pi.
$$
We refer to their paper \cite{AR} for the details of the above
scheme of proof.

It is also possible to estimate the rate of decay towards the
equilibrium for the solutions. In fact, if $g_0 \in L^\infty$, there
exist positive constants
$\sigma$ and $k$ such that (see \cite{AR})
$$
\left\|g^{PM}(\cdot,t)  - {\int_{-\pi}^{\pi} g_{0}(x) \, dx \over
2\pi}
\right\|_{L^\infty(-\pi,\pi)} \ \leq \ k e^{-\sigma\, t}\,.
$$

We now come back to the complete hard-needles problem for nonzero
temperature. The splitting method consists of solving in $[0, \Delta
t]$ the PME ($T=0$) and diffuse it by the heat kernel ($T>0$) to
obtain $g_1 = g^S(x,\Delta t) =
\Theta(\Delta t,x) \ast_x U(\Delta t) g_0 $, where $g^S$ means the
splitting solution and $\ast_x$ is
$$
\Theta(x,\Delta t) \ast_x f(x) = \int_{-\pi}^{\pi} \Theta(x-y,\Delta
t) f(y) \ dy.
$$
The method is well defined and follows the same scheme in every
interval $[n\Delta t, (n-1)\Delta t]$. The iterative map $g_n \to
U(\Delta t) g_{n-1} \to \Theta \ast_x U (\Delta t) g_{n-1}$ is a
composition of contractive maps in $L^1$. The diffusivity of
$\Theta$ implies $g > 0$, for $t>0$, and the nonexistence of a
propagation front, i.e. of a free boundary. Also the splitting
argument shows the convergence in
$L^1$ of the solutions of the hard-needles equation to the PME as $T
\to 0$.


\subsection{The stick-needles model}
The dynamical equation for the
stick-needles model is
\begin{equation}
{\partial g \over \partial t} \ = \ - K_0{\partial
\over \partial x} \left(g {\partial g \over \partial x} \right) \ +
\ T{\partial^2 g \over \partial x^2}.
\label{PM-}
\end{equation}

An analysis similar to the hard-needles model may be done in the
present case. However, there is a hurdle to be resolved previously:
the model is not well-possed unless (see \cite{LSU})
$$
g \leq {T \over K_0}.
$$
Moreover this condition implies the non convergence to a
well-possed problem as $T \to 0$ because the generating function
must be non negative.



\section{The Burgers model}
\label{sec-burgers}
\begin{equation}
{\partial g \over \partial t} \ = \ 2K_0 g
{\partial g \over \partial x} \ + \ T{\partial^2 g \over \partial
x^2}
\label{B1}
\end{equation}

The existence and uniqueness properties of this
equation are well understood, see \cite{GR} for
the main results and references. In \cite{Pa}, the
Hopf-Cole transformation is used to analyze the
Burgers equation with periodic boundary
conditions.

To deduce some some decay estimates for small
initial data in $L^1_{2\pi}$, we can repeat the
same arguments as in the study of the
hard-needles model by using the integral
formulation
\begin{eqnarray}
g(x,t) \ = \ \int_{-\pi}^{\pi}\Theta(x-z,t)
g_0(z)\ dz \nonumber\\
+ {1 \over 2}\int_{-\pi}^{\pi}\int_0^t {\partial
\Theta(x-z,t-s) \over \partial z}   g^2(z,s) \, dz ds.
\label{i2}
\end{eqnarray}
This procedure leads, for $1 < p \leq \infty$, to
\begin{equation}
\| g(\cdot,t)   - {1\over 2 \pi}\, \int_{-\pi}^{\pi} g_0(x) \, dx
\|_{L^p_{2\pi}}
\leq c t^{- \gamma} \| g_0
\|_{L^1_{2\pi}}.
\label{BLp}
\end{equation}

\begin{figure}
\begin{center}
\leavevmode
\epsfysize=300pt{\epsffile{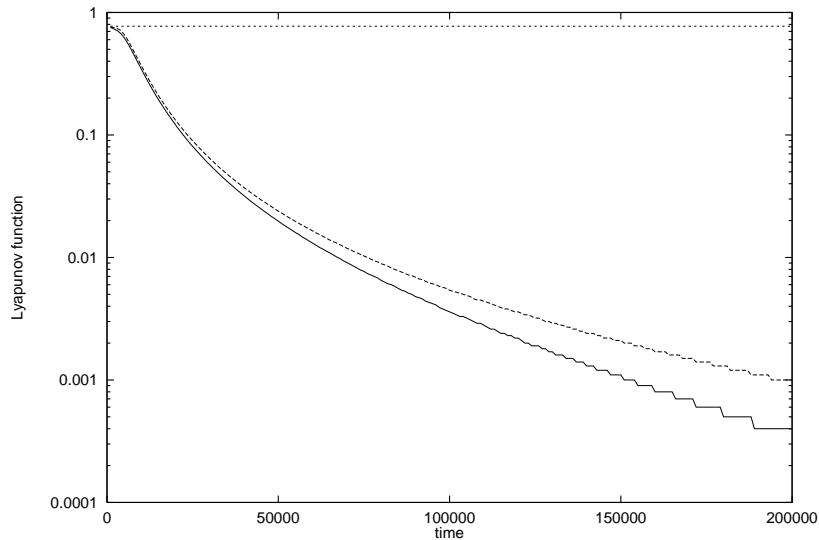}}
\end{center}
\protect\caption[3]{Time evolution of the
Liapunov function for the following values of
temperature: a) T=0 (dotted line), b) T=0.05
(dashed line), and c) T=0.1 (solid line).
\protect\label{flia}}
\end{figure}

In the case of the Burgers equation it is also possible, as in the
Porous Medium case, to deduce that
$$
G(t) = {1\over 2}\,\int_{-\pi}^{\pi} g(x,t)^2\, dx
$$
is a Liapunov functional only if
$T>0$  and its derivative satisfies
\begin{equation}
G'(t) = - T \int_{-\pi}^{\pi} \left({\partial
g\over\partial x}\right)^2 dx
\leq 0.\label{lia3}
\end{equation}
Then, repeating the same ideas as in Section
\ref{sec-odd}, we can prove that  $g(x,t)$ should evolve towards
$(2\pi)^{-1}$ if $T>0$, after normalization. At zero temperature
$G(t)$ is an integral of motion and the previous argument cannot
be used. Figure\ref{flia} shows the time evolution of $G(t)$ for
different temperatures. For $T=0$ it remains constant while for
$T>0$ it decays towards the expected value.


On the other hand, the spatially periodic solution of the Burgers
equation converges towards an entropy solution of the Hopf equation
\begin{eqnarray}
{\partial g^H \over \partial t} \ = \ 2K_0 g^H
{\partial g^H \over \partial x},  \label{H1}\\ g^H(-\pi,t) \ = \
g^H(\pi,t), \label{ccH}\\ g^H(0,x) \ = \ g_0(x) ,
\label{ciH}
\end{eqnarray}
as $T \to 0$. This follows by using same arguments as
in the hard-needles model. In \cite{Ou}, K.\ L.\ OuYoung proved the
following result based upon ideas developed by M. G. Crandrall for
the Cauchy problem in $\RR$: the semigroup $S: g_0 \to S(t) g_0=
g^H$ is a contracting semigroup in
$L^1_{2\pi}$ and we have
\begin{eqnarray}
\| g^H(\cdot,t) \|_{L^\infty_{2\pi}} \leq \| g_0
\|_{L^\infty_{2\pi}}, \quad
\mbox{a.e.,}\nonumber\\
 TV( g^H (\cdot,t) ) \leq TV( g_0),\quad\quad\quad\nonumber\\
\int_{\pRR} |g^H (t_2, x) - g^H (t_1, x) | \ dx \leq cTV( g_0)|t_2 -
t_1|,\nonumber
\end{eqnarray}
for $t_1, \ t_2 \geq 0$. The uniqueness of solution
for (\ref{H1}) is based in the fundamental work of S. N. Kruzhkov,
\cite{K}.

Then, the splitting argument used in the hard-needles model give the
convergence in $L^1_{2\pi}$ for the solution of the Burgers equation
to the entropy solution of the Hopf equation.

\begin{figure}
\begin{center}
\leavevmode
\epsfysize=300pt{\epsffile{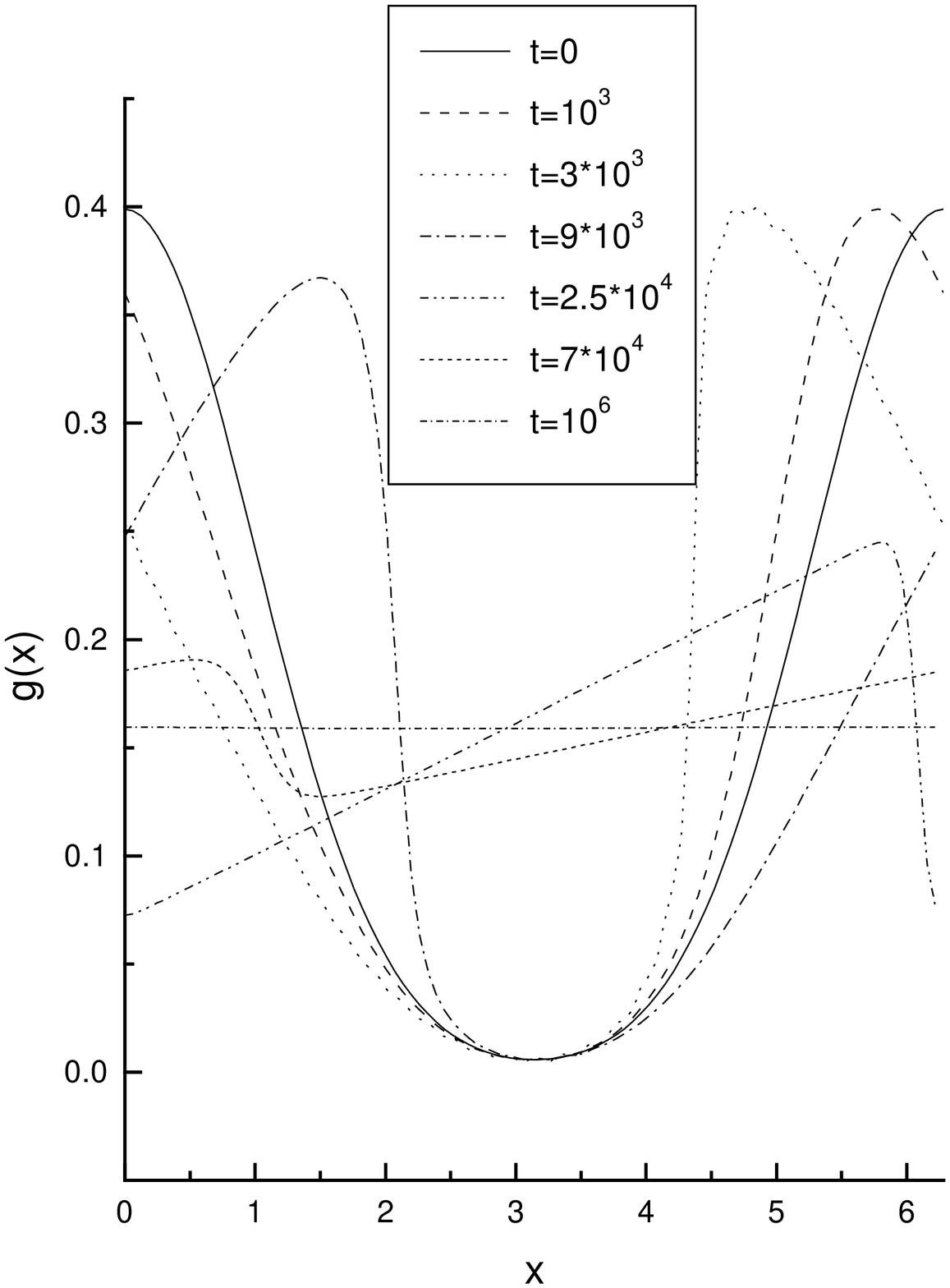}}
\end{center}
  \protect\caption[3]{Seven snapshots of the generating function $g$ taken
at different
times. We have chosen as initial state a Gaussian distribution.
Notice the appearance of the shock wave.\protect\label{shock}}
\end{figure}

It is well-known that in general (\ref{H1}), (\ref{ccH}) and
(\ref{ciH}) do not have continuous solutions for all time because
shock waves appear after finite time and tend to dominate the
solution as $t\to\infty$. The quantities $\int_{-\pi}^{\pi} g^k dx$ are
preserved in time until a shock is formed. After a shock wave appears, only
the mass $\int_{-\pi}^{\pi} g \; dx$ is conserved along the time evolution.
Figure \ref{shock} is an illustration
of the phenomenon.  E.\ Hopf in \cite{Ho} discussed the asymptotic
behaviour for large time of solutions of (\ref{H1})-(\ref{ccH}).
Later, P. D. Lax \cite{L} and T. P. Liu and M. Pierre \cite{LP}
completed and generalized this study for an arbitrary conservation
law. A first result in this direction is given by the following
inequality
\begin{equation}
|g^H(x,t) -M(x,t)| \ \leq \ c{2\pi \over t},
\label{ab1}
\end{equation}
where $M(x,t)$ is the mean function
$$
M(x,t) \ = \ {1 \over 2\pi} \int_{-\pi}^{\pi} g^H(x,t) \ dx .
$$
If $g_0$ is in $L^\infty_{2\pi}$, then the mean function is time
independent and (\ref{ab1}) holds.

Let us assume that initially $g_0(x)$ has a single maximum and
a single minimum at each period. Then $g^H$ evolves towards a
precise asymptotic shape so that its profile for large time looks
like a 2$\pi$-periodic sawtooth profile with a single shock per
period. Between  shocks $g^H$ decreases linearly. The asymptotic
behaviour profile result in the case of a Cauchy problem was
specified by  T. P. Liu and M. Pierre in \cite{LP}. By adapting
the arguments in \cite{LP}, we can prove for our periodic problem

$$
\lim_{t \to \infty} t^{1 \over 2p'} \|g^H(\cdot,t) - \omega(\cdot,t)
\|_{L^p_{2\pi}} \ = \ 0,
$$
for an initial data $g_0 \in L^1_{2\pi}$ and $1 \leq p < \infty$.
Here

\begin{eqnarray}
\omega\left(x,t\right) \, = \,
{1 \over 2 \pi} \, - \, {x\over 2K_{0}t} \, =\, -{1 \over 2 K_0 t}
\left( x - {K_0 \over \pi} t \right) \qquad -\pi < x < \pi
\end{eqnarray}
which is a sawtooth profile moving towards the right at speed
$K_0/\pi$. The strength of the shock $\mbox{max}(g) -\mbox{min}(g)
= \pi/(K_0 t)$ converges to zero as the time goes to infinity.
Thus the shock weakens, it vanishes and we are left with the
incoherent solution $(2\pi)^{-1}$. In summary, incoherence is the
globally asymptotically stable solution for the periodic Burgers
equation at any temperature, including zero. In the last case,
interesting transient patterns including shock profiles are
possible before every memory of the initial conditions is washed
out.

\section{The extended Burgers model}
\label{sec-daido} Instead of working with the nonlocal model given
by Equations (\ref{eq8}) and (\ref{v(x,t)}) for $g(x,t)$ we can try
to find local equations for $v(x,t)$ and $g(x,t)$. Let us define
\begin{eqnarray}
\rho(x,t) = 2K_{0}\, [g(x,t) - g(x+\pi,t)],\nonumber\\
\sigma(x,t) = 2K_{0}\, [g(x,t) + g(x+\pi,t)] .\label{definitions}
\end{eqnarray}
These functions obey the equations (\ref{v(x,t)}) and
\begin{eqnarray}
\frac{\partial \rho}{\partial t} = - \frac{\partial (\sigma
v)}{\partial x} + T \frac{\partial^2 \rho}{\partial
x^2}\,,\label{rho}\\
\frac{\partial \sigma}{\partial t} = \frac{\partial}{\partial x}
\Bigl (v\frac{\partial v}{\partial x}\Bigr ) + T \frac{\partial^2
\sigma}{\partial x^2}\,.\label{sigma}
\end{eqnarray}
Now we can obtain an equation for $v(x,t)$ by
inserting (\ref{dV}), i.e.
$\partial v/\partial x = -\rho$ in (\ref{rho}), and integrating the
result  with respect to $x$:
\begin{eqnarray}
\frac{\partial v}{\partial t} - \sigma v - T \frac{\partial^2
v}{\partial x^2} = J(t)\,.\label{SMeq}
\end{eqnarray}
Equations (\ref{v(x,t)}) and (\ref{definitions})
together with the periodicity of $g(x,t)$ imply that
\begin{eqnarray}
\sigma(x+\pi,t) = \sigma(x,t), \quad v(x+\pi,t) = - v(x,t)
,\label{SMc}
\end{eqnarray}
i.e. the $2\pi$-periodic functions $\sigma(\cdot,t)$
and $v(\cdot,t)$ are periodic and antiperiodic in the $x$ variable,
respectively. Moreover the normalization condition for $g$ and the
$\pi$-periodicity of $\sigma$ imply
\begin{eqnarray}
\int_0^{\pi} \sigma(x,t) \ dx = 2K_{0} .\label{normalization}
\end{eqnarray}
The conditions (\ref{SMc}) may be used to show that
$J(t)\equiv 0$ in (\ref{SMeq}), so that $v$ obeys
\begin{eqnarray}
\frac{\partial v}{\partial t} - \sigma v - T \frac{\partial^2
v}{\partial x^2} = 0\,.\label{SM}
\end{eqnarray}
To see this, substitute $x+\pi$ as the argument of
$v$ and $\sigma$ in (\ref{SMeq}), and use (\ref{SMc}) to obtain
$J=-J$. Once $\sigma$ and $v$ are found, we obtain the generating
function from the equation
\begin{eqnarray}
g(x,t) = {1\over 4 K_{0}} \left[\sigma(x,t) -
{\partial v(x,t)\over\partial x}
\right]. \label{eq.g}
\end{eqnarray}

A similar development to the study of the existence properties of
solutions given for the Porous Medium and Burgers models can be done
for
$v(x,t)$ in  (\ref{SM}). In fact, we can write
\begin{eqnarray} v(x,t) = \int_{-\pi}^{\pi}\Theta(x-z,t)  v_0(z)\ dz
\nonumber \\ +  \int_{-\pi}^{\pi}\int_0^t  \Theta(x-z,t-s)
\left( \sigma \, v \right) (z,s) \ dz \ ds
\label{v1}
\end{eqnarray}
coupling it with
\begin{eqnarray}
\sigma(x,t) = \int_{-\pi}^{\pi}\Theta(x-z,t)  \sigma_0(z)\
dz\quad\quad\quad\quad\quad\quad
\quad\quad \nonumber\\ +  \int_{-\pi}^{\pi}\int_0^t {\partial
\Theta(x-z,t-s) \over \partial z}
\left( v {\partial v \over \partial z} \right) (z,s) \ dz \ ds.
\label{s1}
\end{eqnarray}
The same type of estimates as in the previous
sections together with an iterative method and a fix point theorem
lead to a cubic equation, instead of a quadratic one as in
(\ref{M}). This allows to ensure the existence and  bounds in
$L^p$, at least for small initial data in $L^1$, of both $v(x,t)$
and $\sigma(x,t)$, for positive temperature $T$. However,  the
Theta function does not appear with a space derivative in
(\ref{v1}) as in the Porous Medium and Burgers models which
implies the non-convergence in time towards its respective mean
values.

For this extended Burgers model the asymptotic in time behaviour
is described by the Liapunov functional analyzed in Section
\ref{sec-odd}. Let us now find the stationary solutions of these
equations. From (\ref{sigma}) and (\ref{SM}), we obtain
\begin{eqnarray}
\sigma v + T \frac{d^2 v}{d x^2} = 0 ,\label{st-sigma}\\ T\sigma +
{v^{2}\over 2} = L + M x . \nonumber
\end{eqnarray}
$\sigma(x)$ and $v(x)$ are $2\pi$-periodic functions obeying
(\ref{SMc}) and
$L$ and $M$ are constants. $2\pi$-periodicity of $\sigma$ and
$v$ imply that $M=0$ in the second of these expressions. Thus
\begin{eqnarray}
T\sigma + {v^{2}\over 2} = L . \label{L}
\end{eqnarray}
Now we have $L> 0$ because $\sigma > 0$. Combining
(\ref{normalization}) and  (\ref{L}), we find the following equation
for $L$:
\begin{eqnarray}
2TK_0 + \int_0^\pi {v^{2}\over 2}\, dx = \pi L .
\label{L.int}
\end{eqnarray}

Inserting (\ref{L}) into  (\ref{st-sigma}) we obtain
\begin{eqnarray}
T^2 \, \frac{d^2 v}{dx^2} + \left( L - {v^{2}\over
2}\right)\, v = 0.\label{eB1}
\end{eqnarray}
This may be integrated once yielding
\begin{eqnarray}
{1\over 2}\, \left(\frac{dv}{dx}\right)^2 + {L
v^{2}\over 2 T^{2}} - {v^{4}\over 8T^{2}} = {E\over
T^{2}}\,.\label{eB2}
\end{eqnarray}
The energy $E$, $0\leq E< L^2/2$ is calculated so
that the period of
$v(x)$ be $2\pi$. It is convenient to rewrite the previous equation
in terms of
\begin{eqnarray}
w(\xi) = {v(x)\over \sqrt{L}}\,,\quad x = {T\xi
\over \sqrt{L}}\,.\label{eB3}
\end{eqnarray}
Then (\ref{eB2}) becomes
\begin{eqnarray}
{1\over 2}\, \left(\frac{dw}{d\xi}\right)^2 +
{w^{2}\over 2} - {w^{4}\over 8} = {E\over L^{2}}\equiv {\cal
E}.\label{eB4}
\end{eqnarray}
Suppose that we choose $w(0)=0$, $w'(0)>0$ as initial
conditions for a given trajectory. Then $w(\xi)$ is
\begin{eqnarray}
\int_0^{w(\xi)} {dw\over \sqrt{2{\cal E} - w^{2} + {w^{4}\over 4}}}
= \xi,\label{eB5}
\end{eqnarray}
until $v$ reaches the turning point $w_0({\cal E})$,
where
\begin{eqnarray}
w_{0}({\cal E}) = \sqrt{2}\, \sqrt{1-
\sqrt{1-2{\cal E}}}\,,\label{eB6}
\end{eqnarray}
at $\xi = P({\cal E})/4$. Notice that we will obtain
a one-parameter  family of stationary solutions because $w(\xi +
c)$, with $w(\xi)$ as in  (\ref{eB5}) and $c \in \RR_+$, is also an
admissible stationary solution.  Because of symmetry, an oscillation
period is first completed at four times  this value, i.e., when
\begin{eqnarray}
 P({\cal E}) = 4 \int_0^{w_{0}({\cal E})} {dw\over
\sqrt{2 {\cal E} -
w^{2} + {w^{4}\over 4}}}  = {4\sqrt{2}\,
K(m)\over\sqrt{1+\sqrt{1-2{\cal E}}}}\,, \label{eB7}\\ m =
{1-\sqrt{1-2{\cal E}}\over 1+\sqrt{1-2{\cal E}}}\,.\label{eB7.1}
\end{eqnarray}
Here $K(m)$ is the complete elliptic integral of the
first kind with  parameter $m$ given by (\ref{eB7.1}); see
\cite{AS}, page 590. $P(\cal{E})$  is an increasing function from
$P(0)=2\pi$ to $P(1/2) = +\infty$. Branches  of admissible
stationary solutions satisfy (\ref{SMc}), in particular they
fulfil $v(x+\pi) = - v(x)$. Then their energies ${\cal E}\in
(0,1/2)$ should  be such that
\begin{eqnarray}
n\, P({\cal E})  = {2\pi\sqrt{L}\over
T}\,,\label{eB8}
\end{eqnarray}
for positive odd integers $n=2p-1$, $p=1,2,\ldots$.
Notice that even $n$'s  would yield inadmissible solutions such that
$v(x+\pi) = + v(x)$.  By changing to the $w$ and $\xi$ variables,
Eq.\ (\ref{L.int}) for $L$ may  be written as
\begin{eqnarray}
n\, \int_0^{w_{0}({\cal E})} {w^{2} dw\over \sqrt{2
{\cal E} -
w^{2} + {w^{4}\over 4}}}  = {\pi\sqrt{L}\over T} -
{2K_{0}\over\sqrt{L}}
\label{eB9}
\end{eqnarray}
($n$ odd) or, equivalently, as
\begin{eqnarray}
n \sqrt{8}\sqrt{1+\sqrt{1-2{\cal E}}}\, [K(m) -
E(m)]  =  {\pi\sqrt{L}\over T} - {2K_{0}\over\sqrt{L}} .
\label{eB10}
\end{eqnarray}
Here $E(m)$ is the complete elliptic integral of the
second kind with  parameter $m$ given by (\ref{eB7.1}).

Given $K_0>0$ fixed, Equations (\ref{eB7}), (\ref{eB8}) and
(\ref{eB10}) yield  the values of ${\cal E}$ and $L$ corresponding
to a branch of stationary  solutions with a given odd value of $n$.
A given stationary solution
$(n,{\cal E},L)$ will have $n$ maxima and $n$ minima. We can find
$L$ from  (\ref{eB8}) and eliminate it in (\ref{eB10}) with the
result
\begin{eqnarray} {4 n^{2}\, K(m)\over\pi}\, \left[ E(m)
 - {\sqrt{1-2{\cal E}}\, K(m)\over 1+\sqrt{1-2{\cal E}}}\right]  =
{K_{0}\over T}\, . \label{eB11}
\end{eqnarray}
This latter equation relates the energy ${\cal E}$ to
the coupling parameter
$K_0$. As $K_0$ increases so does the number of possible stationary
branches.  Notice that the stationary branch with index $n$ exists
for values of the  coupling constant larger than $K_0 = n^2 \,
T\pi/2$. At this value, ${\cal E}  =0$ (therefore, $m=0$). Clearly
the number of possible stationary branches is  then half the integer
part of $\sqrt{2K_{0}/(\pi T)}$ (only odd $n$ are  admissible),
which increases with $K_0$. For a given stationary solution
$v(x)$, the stationary probability density may be reconstructed from
(\ref{eq.g}) as
\begin{eqnarray}
g(x) = {1\over 4K_{0}}\,\left\{ {L\over T}
-{v^{2}\over 2T}  - {dv\over dx} \right\}\,. \label{gstat}
\end{eqnarray}

A stationary periodic moment-generating function is given by
(\ref{eB2}),  (\ref{eB3}), (\ref{eB5}), (\ref{eB8}), (\ref{eB9}),
(\ref{eB11}) and  (\ref{gstat}). Their explicit form may be obtained
by using the definitions  of the Jacobi elliptic functions; see
\cite{AS}, pages 569 and ss. From  (\ref{eB3}), (\ref{eB5}),
(\ref{eB8}), (\ref{eB11}) and (\ref{OF}), we find
\begin{eqnarray}
v(x) = {4nT\sqrt{m} K(m)\over\pi}\, \mbox{sn} u\,,
\label{eB12}\\ H(x) = -{n^{-1}\, m^{{1\over 2}}\over E(m) -
{\sqrt{1-2{\cal E}}\, K(m)\over  1+\sqrt{1-2{\cal E}}} }\, \mbox{sn}
u\,,
\label{OFstat}\\ u = {2n K(m) x\over\pi}\,.\label{eB13}
\end{eqnarray}
The probability density may be reconstructed from
(\ref{eB8}), (\ref{eB11})  -- (\ref{eB13}):
\begin{eqnarray}
g(x) = {K(m)\over 2\pi}\, { {1\over
1+\sqrt{1-2{\cal E}}} -  m\, \mbox{sn}^2 u - \sqrt{m}\,\mbox{cn}
u\,\mbox{dn} u\over E(m)  - {\sqrt{1-2{\cal E}}\, K(m)\over
1+\sqrt{1-2{\cal E}}}}\,.
\label{eB14}
\end{eqnarray}
Notice that these synchronized solutions are defined
up to a constant phase  shift $x\to x+c$, as we said before. Such
solutions exist for any positive  temperature $T>0$ if $K_0\geq
n^2 T\pi/2$.  Therefore, it is possible to find synchronized
solutions of the extended  Burgers model for any positive
temperature. The stationary drift velocity  (\ref{eB12}) has
$n=2p-1$ ($p=1,2,\ldots$) maxima and $n$ minima on the interval
$-\pi<x<\pi$. Between successive extrema, $v(x)$ vanishes once.
According to  (\ref{gstat}), the extrema of $g(x)$ on the interval
$(-\pi,\pi]$ are reached at  the zeros of $v(x)$. A little algebra
shows that
$$
{dg\over dx} = {L^{{3\over 2}}\over 4K_{0}T^{2}}\,
w(\xi)\, \left[1-{w^{2} \over 2} -{dw\over d\xi}\right]\,,
$$
and thus
$$
{d^{2}g\over dx^{2}}\Big|_{v=0} = {L^{2}\over
4K_{0}T^{3}}\, \left[{dw\over d\xi} - \left({dw\over
d\xi}\right)^{2}\right]\,.
$$
Since $dw/d\xi = \pm\sqrt{2{\cal
E}}$ wherever $w=0$, we have
$$
\mbox{sign}\left({d^{2}g\over
dx^{2}}\right)\Big|_{v=0} = \mbox{sign}\left({dw\over
d\xi}\right)\,.
$$
Therefore maxima of $g(x)$ are reached at
points where $v=0$ and $v'(x)<0$.  These zeros are $u = 2 l K(m)$,
where $l$ is an odd number running from $-n+2$ to $n$. The
corresponding values of $x$ are $x_{l,n} = l\pi/n$. Thus  an
oscillator population described by (\ref{eB12}) is split in $n$
subpopulations with angles close to $x=x_{l,n}$,
$l=-n+2,\ldots,n$. The  frequency density is found by means of
(\ref{limit9}). In dimensionless form, $p_n(\Omega) = \sqrt{L}\,
P_n(\sqrt{L}\Omega)$, which yields
\begin{eqnarray}
p_n(\Omega) & = & {nTK(m) w_{0}({\cal E})\over 2\pi
K_{0}\sqrt{2 {\cal E}}}
 \sum_{w(\xi) = \Omega} \left\{ {1-{\Omega^{2}\over 2}\over
|w'(\xi)|} - \mbox{sign}(w'(\xi)) \right\}\,.
\label{p_n}
\end{eqnarray}
Here the sum is over all solutions $\xi\in
(-\pi,\pi]$ of the equation $w(\xi)  = \Omega$. The maxima of
$p_n(\Omega)$ yield the likeliest frequencies of  rotation for the
oscillators. See \cite{daidoOF,daido} for additional physical
interpretation and numerical simulations.

\begin{figure}
\begin{center}
\leavevmode
\epsfysize=200pt{\epsffile{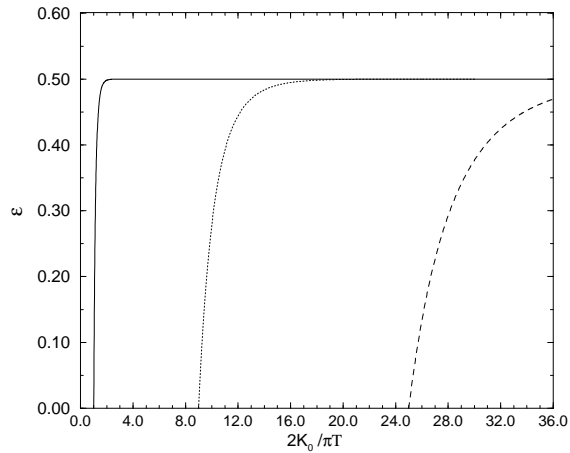}}
\end{center}
  \protect\caption[2]{Bifurcation diagram
${\cal E}$ versus $2K_{0}/(\pi
T)$ for stationary synchronized solutions branching off from incoherence
${\cal E} = 0$ at the square of the odd integer numbers.
\protect\label{FIG2} }
\end{figure}

Once a stationary solution (with index $n$), $g(x)$ [equivalently we
may  specify $\sigma(x)$ and $v(x)$], has been found, it is
interesting to discuss  its linear stability. Let $\sigma(x,t) =
\sigma(x)+ e^{\lambda t}
\hat{\sigma}(x)$, $v(x,t) = v(x)+ e^{\lambda t} \hat{v}(x)$. Then
$\hat{\sigma}$ and $\hat{v}$ satisfy (\ref{SMc}), $\int_0^{\pi}
\hat{\sigma} dx = 0$, and solve:
\begin{eqnarray}
 - T \frac{d^2 \hat{v}}{dx^2} - [\sigma(x) \hat{v} +
v(x)\hat{\sigma}] = -\lambda \hat{v}\,, \label{stab1}\\
 - T \frac{d^2 \hat{\sigma}}{dx^2} - \frac{d^2 [v(x)\hat{v}]}{dx^2}
= -
\lambda \hat{\sigma} \,. \label{stab2}
\end{eqnarray}
Solving these equations in the general case seems
difficult. These equations are easy to solve for
the incoherent solution $v(x)=0$,
$\sigma(x) = 2K_0/\pi$. In this case (\ref{stab1}) and (\ref{stab2})
uncouple and we explicitly find the eigenvalues
\begin{eqnarray}
\lambda = {2K_{0}\over\pi} - (2p-1)^2 T,\label{stab3}\\
\lambda = - (2p)^2 T,\label{stab4}
\end{eqnarray}
(where $p = 1,2,\ldots$) corresponding to
(\ref{stab1}) and (\ref{stab2}), respectively. Eq.\ (\ref{stab3})
shows that incoherence is linearly unstable for $K_0 > T \pi/2$.  At
$K_0 = (2p-1)^2 T \pi/2$ the stationary solutions (\ref{eB14}) (with
$n=1,3,\ldots$) bifurcate from the incoherent solution.  This can also
be checked as follows: linearize (\ref{eB1}) about $v=0$ and calculate
the value of $K_0$ for which there is a $2\pi$-periodic solution of
the resulting equation which satisfies (\ref{SMc}). It is $K_0 = T
\pi/2$ corresponding to $E=0$, $n=1$. This solution is linearly stable
(at least for values of $K_0$ in an appropriately small
half-neighbourhood of $T \pi/2$) because of the principle of exchange
of stabilities between bifurcation branches. In
figure \ref{FIG2} we show the different
bifurcation branches which occur at different
integer values of $n=2p-1$.

\begin{figure}
\begin{center}
{\epsfysize=230pt
\epsffile{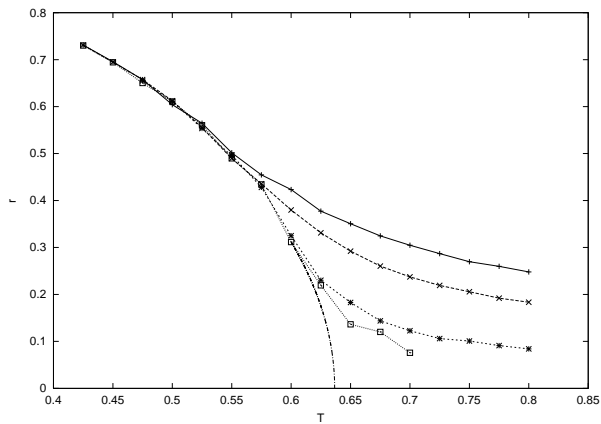}}

  \protect\caption[2]{Synchronization parameter
$r$ as a function of $T$ for different sizes
$N=50,100,500,1000,\infty$.
\protect\label{FIG3} }
\end{center}
\end{figure}

An alternative calculation of the first branch
(index $n=1$) of stationary solutions bifurcating
from incoherence is using Monte Carlo methods.
They are quite powerful since acceptance rate can
be tuned at will in such a way that the approach
to the stationary state may be faster.
Furthermore, the intrinsic dynamics in Monte
Carlo methods is discrete in time, unlike the
continuous-time dynamics of the original
problem. Thus the rounding errors due to time
discretization are absent in Monte Carlo
calculations of stationary solutions. We use the
Monte Carlo method and the Glauber algorithm with
a random sequential updating of the phases for
the model Hamiltonian (\ref{eq13}) of Section
\ref{sec-family}.(d). Local phases are randomly
changed from $\phi_i$ to $\phi_i+\alpha\delta$,
where $\alpha$ takes values with equal
probability on the interval $[0,1]$ and
$\delta$ is the typical size of the move. The
proposed change is accepted with probability
$[1+\exp(\beta \Delta {\cal H})]^{-1}$ where
$\Delta {\cal H}$ is the change in the
Hamiltonian.  A convenient synchronization order
parameter $r$ is defined through the global
magnetization $M=r e^{i\alpha}=(1/N)\sum_{j=1}^N
e^{i\phi_{j}}$. The order parameter can be
calculated from Monte Carlo simulations of the
Hamiltonian (\ref{f1}) with $E(x)=|x|$ and
$K_0=1$; see figure \ref{FIG3}. Notice that the
curves in this figure tend to a curve
intersecting the horizontal axis at the
bifurcation point $T=2/\pi$ as $N$ increases.
This corresponds to the expected bifurcation
result for the NLFPE (dot-dashed line in Figure
\ref{FIG3} corresponding to $N=\infty$). Notice
that finite-size corrections are of order
$N^{-{1\over 2}}$ far from the bifurcation point
and of order $N^{-{1\over 4}}$ near it
\cite{dawson}.

\begin{figure}
\begin{center}
\leavevmode
\epsfysize=240pt{\epsffile{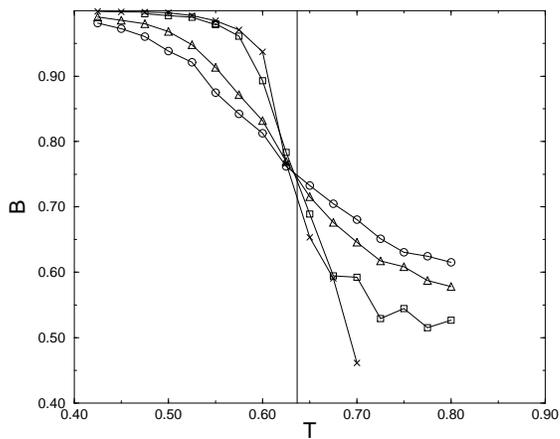}}
\end{center}
  \protect\caption[2]{Binder parameter $B$ as a
function of $T$ for different sizes (in the low
$T$ region, from bottom to top)
$N=50,100,500,1000$. The curves become steeper
for larger values of $N$. The theoretical value of the bifurcation
temperature $T=2/\pi$ is marked in the figure.
\protect\label{FIG4} }
\end{figure}

 The transition temperature corresponding to the
first branch is easily obtained through standard
finite-size scaling methods. Consider the
kurtosis (or Binder parameter) for the
synchronization parameter $r$, $B =\frac{1}{2}(3
-\frac{<r^4>}{<r^2>^2})$, where $<\ldots>$ is the
standard configurational average [weighted with
the usual Boltzmann-Gibbs factor, $\exp(-\beta
{\cal H})$, and ${\cal H}$ is given by
(\ref{eq13})]. The curves for $B$ are shown in
figure \ref{FIG4} for different sizes.  Note that
these curves (specially for $N=50, 100, 500$,
data for $N=1000$ is more noisy) intersect at a
common point characterizing the bifurcation
temperature.

It is interesting to compare the present results
with those of Daido's order function
theory~\cite{daido}. Except for a constant
factor, $-K_0^{-1}$,  Daido's order function is
just $v(x,t)$ in our notation. Notice that  as
$K_0/T \to\infty$, (\ref{OFstat}) and
(\ref{eB14}) corresponding to  the linearly
stable solution with index $n=1$ become $H(x) = -
$ sign$(x)$  and $g(x) = \delta(x-\pi)$,
respectively. This coincides exactly with
Daido's solution at $T=0$ \cite{Cr,daido}. An
interesting aspect  of our exact construction of
stationary solutions is that they can shed some
light on scaling near bifurcation points as $T\to
0$ \cite{Cr,daido}.  In fact, (\ref{eB11}) implies
that stationary solution branches issue forth
from incoherence at $K_0 = \kappa_n \equiv n^2 \pi
T/2$ ($n$ odd). Furthermore  (\ref{eB11}) shows
that the $n$th synchronized stationary solution
branch satisfies ${\cal E}\propto (K_0/T - n^2
\pi/2)$ for $K_0/T$ close to its bifurcation
value $K_0 = \kappa_n$. Then (\ref{eB12}) -
(\ref{eB14}) show that $g(x) = O(|K_0 -
\kappa_c|^{1/2})$ if $T>0$ ($\kappa_c$ stands
for the value of the bifurcation parameter
$K_0/T$ at the bifurcation point $\kappa_n$).
Clearly as $T\to 0+$, a quasicontinuum of
$N=O(\sqrt{K_{0}/T})$ stationary branches has
bifurcated from incoherence  for a small fixed
$K_0$. In these circumstances, the derivation of
a one-mode amplitude equation as in Ref.\
\cite{Cr} does not describe correctly the
situation. See \cite{BH95} for a derivation of an
amplitude partial differential equation
describing a quasicontinuum of Hopf bifurcations;
similar techniques could be used in the present
case.

The results at zero temperature may be obtained
in another form. The model with Daido coupling
is described by (\ref{sigma}), (\ref{SM}) and
(\ref{eq.g}).  Setting $T=0$ in these equations,
we obtain an integro-differential equation
$$
{\partial v(x,t) \over \partial t} = v(x, t) \left\{ \sigma_0 (x)
\, + \,
\int_0^t {\partial \over \partial x} \left( v(x,s) {\partial v(x,s)
\over
\partial x} \right) \, ds
\right\}
$$
which can be consider as a modification of the Porous Medium
equation with memory. This equation can be also written in a local
form in the following way:
$$
\sigma = {1\over v}\, {\partial v\over\partial t}\,,\quad  {\partial
\sigma\over\partial t} = {1\over 2}\, {\partial^{2} v^{2}\over
\partial x^{2}}\,.
$$
Then $v(x,t)$ obeys
\begin{eqnarray}
{\partial \over\partial t}\left({1\over
v}{\partial v\over\partial t}\right)  = {1\over 2}\, {\partial^{2}
v^{2}\over\partial x^{2}}\,,\label{hyper1}\\ {\partial^{2}
u\over\partial t^{2}} = {1\over 2}\, {\partial^{2} e^{2u}\over
\partial x^{2}}\,,\label{hyper2}
\end{eqnarray}
if $|v| = e^u$. This is an amplitude equation with
infinitely many terms (e.g., expand the exponential in a Taylor
series), and $g(x,t) = O(K_0)$ in agreement with Daido's results:
the critical exponent is $\beta = 1$  if $g=O(|K_0 - K_c|^\beta)$.
Nevertheless, (\ref{hyper1}) or (\ref{hyper2})  are very different
from Crawford's amplitude equations \cite{Cr}.

Clearly the stationary solutions of (\ref{hyper1}) compatible with
the  symmetry requirement $v(x+\pi,t)=-v(x,t)$ are $v_0(x)= q$
sign$(x)$,  where $q$ is a constant. In particular, incoherence
corresponds to
$q=0$. $\sigma_0(x)$ obeys $\sigma v=0$. Thus $\sigma_0$ is zero
except at the points where $v=0$, namely $x=0,\pi$ (mod.\ $2\pi$).
If (\ref{normalization}) holds and $\sigma(x+\pi)=\sigma(x)$, we
have
\begin{eqnarray}
 \sigma_0(x) =  2 K_0 [ \delta(x) +  \delta(x-\pi)] ,
\label{hyperst1}
\end{eqnarray}
on the interval $x\in (-\pi,\pi]$. Inserting
$\partial v_0/\partial x = -2 q\, [\delta(x) - \delta(x-\pi)]$ and
(\ref{hyperst1}) in (\ref{eq.g}),  we obtain
\begin{eqnarray}
g_0(x) = \delta(x-\pi) , \label{hyperst2}
\end{eqnarray}
provided $q=K_0$. This density function is the same
as that obtained above  by taking the limit $K_0/T\to\infty$ of the
stable stationary density with  index $n=1$. Daido found the same
results by using his order function  theory~\cite{daido}. Notice
that the stationary solution (\ref{hyperst2}) is a member of the
one-parameter family of stationary solutions, $g_0(x+c)$,
$c=$ constant.

Linearization of (\ref{sigma}) and (\ref{SM}) (with $T=0$) about
$v_0(x)= K_0$  sign$(x)$ and (\ref{hyperst1}),
$v(x,t) = v_0(x) + V(x,t)$, $\sigma(x,t) = \sigma_0(x) +
\Sigma(x,t)$ ($V\ll v_0$, $\Sigma\ll\sigma_0$), yields
 \begin{eqnarray}
{\partial V\over\partial t} = \sigma_{0}(x)\, V +
v_0(x) \,\Sigma ,
\label{hyperlin1}\\ {\partial\Sigma\over\partial t} = {\partial
v_{0}(x) V\over
\partial x}\,,\label{hyperlin2}
\end{eqnarray}
which is equivalent to
\begin{eqnarray}
{\partial^{2} W\over\partial t^{2}} -
\sigma_0(x)\, {\partial W\over\partial  t} = K_0^2\, {\partial
W\over\partial x} \,,\label{hyperlin3}\\ W(x,t) = v_0(x)\, V(x,t)
.\label{hyperlin4}
\end{eqnarray}
If we prove that $W(x,t)$ evolves towards zero as
$e^{- M t^{2}}$,
$M>0$, then this implies that the density (\ref{hyperst2}) is
linearly stable. But analyzing Equation (\ref{hyperlin3}) as a Heat
equation with respect to
$t$, with boundary conditions $W \to 0$ as $|t| \to \infty$, we have
the Green function of the linear part
$$
\tilde{\Gamma}(x,t) \ = \ (4\pi K_0^{-2}|x|)^{-{1 \over 2}} e^{-
{K_0^2|t|^2
\over 4|x|}}, \ x > 0 .
$$
Using the periodicity and symmetry properties of
$v$ and $\sigma$, the solution can be written in
convolution form with respect to $t$ with initial
data $K_0 V(0, t)$. This allows to assure that
$W(x,t)$ evolves towards zero as $e^{- M t^{2}}$,
with $M= K_0^2/4\pi$.

\section{Concluding remarks}
\label{sec-fin}
In this paper, we have considered the Kuramoto
model for synchronization of phase oscillators
with global periodic coupling functions $f(x)$
and subject to external random forces. For
general couplings and natural frequency
distributions, we have derived the one-oscillator
nonlinear Fokker-Planck equation and given an
approximate formula for its solution in the limit
of high natural frequencies. Provided the
frequency distribution has $m$ peaks in the
high-frequency limit, this formula indicates that
the one-oscillator probability density splits
into $m$ components. Each component corresponds
to the solution of a NLFPE with zero natural
frequency, on a frame rotating with fixed
angular velocity. Which results do we know for
such a reduced zero-frequency NLFPE?

In Section \ref{sec-odd}, we have shown that a
Liapunov functional of the free energy type
exists for a class of odd coupling functions
(zero natural frequency). Then the probability
density evolves towards a stable stationary
function. Which stationary function this is,
depends on $f(x)$. We have analyzed a family of
singular coupling functions for which the NLFPE
reduces to partial differential equations or
systems thereof, such as the porous media
equation, the Burgers equation or systems of
Burgers equations. For these equations, we
have examined the behavior of their solutions
in the limit of large times for different
temperatures; see Section \ref{sec-family} for
details. The most interesting coupling function,
$f(x)=$ sign $x$ (periodically extended outside
$[-\pi,\pi]$), was already studied by Daido in
the case of zero temperature, $T=0$ (see Section
\ref{sec-general}.2 for a rephrasing of his
assumptions and results). This model is capable of
synchronization at any temperature. We have found
exact formulas for stationary synchronized
probability densities which bifurcate
supercritically from incoherence. Although we
have not been able to determine their linear
stability, we know that the first bifurcating
branch of solutions is stable at least for
coupling parameter near the bifurcation point.
Let us assume that this bifurcating branch is
always stable for coupling parameter above the
bifurcation value. We can combine our results to
obtain the stable probability density for a model
with a multimodal natural frequency distribution
in the high-frequency limit. We find that, except
for a constant shift in $x$, the stable
probability density is
\begin{eqnarray}
\rho \sim \sum_{l=1}^{m} \rho^{(0)}_l(x-\Omega_l
\omega_0 t), \label{finaleq1}\\
 \rho^{(0)}_l(x) = \theta\left({\pi T\over 2} -
K_0 \alpha_l \right)\, {1\over 2\pi} + \theta
\left(K_0 \alpha_l - {\pi T\over 2}\right)\,
g(x;K_0 \alpha_l), \label{finaleq2}
\end{eqnarray}
where $\theta(x)$ is the Heaviside unit step
function. In these equations, $g(x;K_0 \alpha_l)$
is the solution (\ref{eB14}) corresponding to
setting $n=1$ and $K_0 \alpha_l$ instead of $K_0$
in (\ref{eB11}). The overall velocity function
(proportional to Daido's order function) is a
superposition of rotating waves correponding to
the contributions of the synchronized components
\begin{eqnarray}
v(x,t) &\sim &
\sum_{l=1}^{m}
\alpha_l\,\theta
\left(K_0 \alpha_l - {\pi T\over 2}\right)\,
{4T\sqrt{m_{l}} K(m_{l})\over \pi}\,\nonumber\\
&\times & \mbox{sn}\left( {2 K(m_{l})\,
(x-\Omega_{l}\omega_{0} t)\over
\pi}\right)\,.
\label{finaleq3}
\end{eqnarray}
Here $m_l$ is the value of $m$ obtained by
substituting $n=1$ and $K_0 \alpha_l$ instead of
$K_0$ in (\ref{eB11}).




\appendix

\section{A convolution identity for the velocity}
\label{app:A}

To obtain the convolution identity (\ref{conv}), we can write the
velocity
$v(x,t)$ as follows
\begin{eqnarray}
v(x,t) & = & -K_0 \sum_{n=-\infty}^{\infty} a_n
H_{-n}^{0} e^{inx}
\nonumber\\ & = &  -K_0 \sum_{n=-\infty}^{\infty} \left( {1 \over
2\pi} \int_{-\pi}^{\pi} e^{-inx'} f(x')  dx' \right) H_{-n}^{0}
e^{inx}
\nonumber\\ & = & -K_0  {1 \over 2\pi} \int_{-\pi}^{\pi} f(x')
\sum_{n=-\infty}^{\infty} e^{in(x-x')}   H_{-n}^{0} \ dx'
\nonumber
\end{eqnarray}
which, using the expression for $g(x,t)$
$$
g(x,t) = {1\over 2\pi}\, \sum_{n=-\infty}^{\infty} H_{n}^{0}
e^{-inx} = {1\over 2\pi}\,\sum_{n=-\infty}^{\infty} H_{-n}^{0}
e^{inx},
$$
gives the announced relationship (\ref{conv}).

\section{Order function and oscillator drift velocity}\label{app:B}
In our notation, Daido's order function is defined as~\cite{Da}
\begin{eqnarray}
H(x,t) = - \sum_{k=-\infty}^{\infty} h_k Z_k
e^{-ikx}\,, \nonumber
\end{eqnarray}
where Daido's coupling function is related to ours by
$h(x) = - f(-x)$ ($h_k = - a_{-k}$), and
\begin{eqnarray}
Z_k(t) =  {1\over N}\,\sum_{j=1}^{N}
e^{ik[\phi_{j}(t)-\Omega_{e}t]} = \int_{-\pi}^{\pi} e^{ikx} \left(
{1\over N}\,\sum_{j=1}^{N}
\delta(\phi_{j}(t)-\Omega_{e}t - x) \right)\, dx \nonumber\\ =
\int_{-\pi}^{\pi} e^{ikx} g(x+\Omega_e t,0,t) dx.\nonumber
\end{eqnarray}
Here we have used that $N^{-1}\sum_{j=1}^{N}
\delta(\phi_j(t) - x)$ tends to
$g(x,0,t)$ in the limit of infinitely many oscillators. Thus
$Z_k(t)$ is $2\pi$  times the Fourier coefficient of $g(x+\Omega_e
t,0,t)$ with index $-k$.  Inserting this in $H(x,t)$ above, we find
\begin{eqnarray}
H(x,t) = - \int_{-\pi}^{\pi} g(x'+\Omega_e t,0,t)
\left(\sum_{k=-\infty}^{\infty}  h_k\, e^{ik(x'-x)}\right) dx'
\nonumber\\ = - \int_{-\pi}^{\pi} g(x'+\Omega_e t,0,t)\, h(x'-x)\,
dx'  = \int_{-\pi}^{\pi} g(x'+\Omega_e t,0,t)\, f(x-x')\,
dx'\nonumber\\ = \int_{-\pi}^{\pi} g(x+\Omega_e t - x',0,t)\,
f(x')\, dx' ,\nonumber
\end{eqnarray}
which, together with (\ref{conv}), implies
$$
H(x,t) = - {1\over K_{0}}\, v(x+\Omega_e t,t) .
$$
Then (\ref{OF}) follows, as said in the Introduction.







{\bf Acknowledgments}
The authors want to express their gratitude to
Rafael Ortega and Juan L. V\'azquez for useful
discussions and for pointing them some
references. We acknowledge partial support by the
DGES (Spain) Projects PB98-0142-C04-01 (LLB) and
PB98-1281 (JS), FOM (The Netherlands) contract
FOM-67596 and DGES (Spain) contract PB97-0971
(FR) and TMR (European Union) contract ERB
FMBX-CT97-0157 (LLB \& JS).

\end{document}